\documentclass[aps,prd,reprint,amsmath,amssymb,nobibnotes,showpacs]{revtex4-1}

\usepackage{graphicx}
\usepackage{dcolumn}
\usepackage{amsfonts}
\usepackage{bm}
\begin{document}
\title{Exploring the effects of a double reconstruction on the growth rate of cosmic structure, using current observational data}
\author{Freddy Cueva Solano}
\affiliation{Instituto de F\'{\i}sica y Matem\'aticas, Universidad Michoacana de San Nicol\'as de Hidalgo\\
Edificio C-3, Ciudad Universitaria, CP. 58040, Morelia, Michoac\'an, M\'exico.}
\email{freddy@ifm.umich.mx,\;\;freddycuevasolano$2009$@gmail.com}
\begin{abstract}
Based on General Relativity (GR) we consider two different cosmological scenarios in where reconstruct the energy exchange ($\bar{Q}$) between cold dark matter ($DM$) fluid 
and dark energy ($DE$) fluid, which is modelled with a $DE$ varying equation of state (EoS) parameter $\omega$. We here investigate the main cosmological effects on the 
growth rate of matter density perturbations ($f\sigma_{8}$), on the effective Hubble friction term ($H_{eff}$), on the effective Newton constant ($G_{eff}$) and on the 
growth index of the linear matter fluctuations ($\gamma$). Our study demonstrates that in the coupled models the evolution of these quantities are modified with respect to 
the predictions in the uncoupled models, and therefore could be used to distinguish among coupled $DE$ scenarios. Finally, we also perform a combined statistical analysis 
using current observational data (geometric and dynamical probes) to put more stringent constraints on the parameters space of the cosmic scenarios studied.
\end{abstract}
\maketitle
\section{Introduction}
The combined statistical analysis of the most recent measurements coming from JLA (Joint Light Curve Analysis) type Ia Supernovae (SNe Ia) data 
\cite{Conley2011,Jonsson2010,Betoule2014}, the growth rate of structure formation obtained from redshift space distortion (RSD) data  
\cite{Jackson1972,Kaiser1987,Mehrabi2015,Alcock1979,Seo2008,Battye2015,Samushia2014,Hudson2013,Beutler2012,Feix2015,Percival2004,Song2009,Tegmark2006,Guzzo2008, 
Samushia2012,Blake2011,Tojeiro2012,Reid2012,delaTorre2013,Planck2015}, the different Baryon Acoustic Oscillation (BAO) 
detected in the galaxy clustering observations ($6$dFGS, SDSS DR $7$, SDSS DR $9$, SDSS DR $11$, BOSS DR $9$ CMASS, $2$dsPCF, $2$dMPS, BOSS DR $11$ CMASS), 
\cite{Blake2011,Planck2015,Hinshaw2013,Beutler2011,Ross2015,Percival2010,Kazin2010,Padmanabhan2012,Chuang2013a,Chuang2013b,Anderson2014a,Kazin2014,Debulac2015,FontRibera2014,
Eisenstein1998,Eisenstein2005,Hemantha2014}, the observations of anisotropies in the power spectrum of the Cosmic Microwave Background (CMB: distance priors) data from the 
Planck $2015$ data \cite{Planck2015,Hu-Sugiyama1996,Bond-Tegmark1997,Neveu2016}, and the Hubble parameter (H) measurements obtained from galaxy surveys 
\cite{Sharov2015,Zhang2014,Simon2005,Moresco2012,Gastanaga2009,Oka2014,Blake2012,Stern2010,Moresco2015,Busca2013} indicate that the present universe is undergoing a phase 
of accelerated expansion. From the theoretical point of view, this phenomenon can be explained introducing in the universe an unknown physical fluid with negative pressure 
so-called $DE$ \cite{Peebles1988,Peebles2003,Sahni2004,Copeland2006}. Many alternative models have been suggested to explain it; in particular, the Lambda Cold Dark Matter 
($\Lambda$CDM) model has a cosmological constant as $DE$ with an EoS parameter $\omega=-1.0$ \cite{Weinberg1989,Sahni2000,Seljak2005,Rozo2010}. However, other more 
structured models replace $\Lambda$ by a dynamical $DE$ such as phantom model \cite{Caldwell2002}, quinton model \cite{Feng2005}, quintessence model \cite{Ratra1988}, 
K-essence model \cite{Picon-Chiba}, Chaplyging gas model \cite{Pasquier-Harko}, massive scalar field model \cite{Garousi-Sami}, etc.\\
In the same way, within the universe we also assume the existence of another dark component so-called $DM$, which acts exactly like the ordinary matter (pressureless), and 
can interact with $DE$ gravitationally.\\
On the other hand, the $\Lambda$CDM model presents two different problems such as the fine tuning and the cosmic coincidence. Then, one way to solve the last problem within 
GR is to assume a coupling between $DE$ fluid and $DM$ fluid. Currently, there are not neither physical arguments nor recent observations to exclude an energy exchange 
between these dark components because their natures are still unknown
\cite{Turner1983,Malik2003,Cen2003,Guo2007,He2008,Bohmer2008,valiviita2008,campo2009,cabral2009,cabral2010,chimento2010,abramo1,Cai-Su,abramo2,cao2011,LiZhang2011}.\\
Due to the absence of a fundamental theory to construct $\bar{Q}$, different phenomenological parameterizations for $\bar{Q}$ have been proposed by mathematical simplicity 
\cite{Cueva-Nucamendi2012}; for example, $\bar{Q}\propto \bar{H}{\bar{\rho}}_{DM}$ \cite{Guo2007,He2008,Bohmer2008,valiviita2008,Cueva-Nucamendi2012}, 
$\bar{Q}\propto \bar{H}{\bar{\rho}}_{DE}$ \cite{Pavon2007,He2008,valiviita2008}, $\bar{Q}\propto{\bar{H}}({\bar{\rho}}_{DM}+{\bar{\rho}}_{DE})$ 
\cite{He2008,Wang2005,Wang2006,Campo2006}, $\bar{Q}\propto{\bar{\rho}}_{DM}$ \cite{Bohmer2008,cabral2009} and $\bar{Q}\propto{\bar{\rho}}_{DE}$ \cite{cabral2009}. 
Then such models may be physically viable, if they are confronted with the observational data, and therefore, can be employed in order to look new physical properties on 
cosmological scales \cite{Zimdahl2005,Das2006,Huey2006,Wang2007}.\\
On the other hand, the properties of $DE$ fluid are mainly characterized by the EoS parameter $\omega$. In this case, two possibilities are proposed to explain a varying 
$\omega$. The first one is to parameterize $\omega$ in terms of some free-parameters \cite{Cooray1999,Chevallier-Linder,Tegmark2004,Barboza2009,Wu2010}. Thus, among all the 
different ansatzes the Chevallier-Polarski-Linder (CPL) parameterization \cite{Chevallier-Linder} $\omega=\omega_{0}+\omega_{1}[z/(1+z)]$ (where $z$ is the redshift, 
$\omega_{0}$ and $\omega_{1}$ are dimensionless parameters) is considered as the most popular ansatz. This ansatz shows a divergence problem when redshift $z$ approaches 
to $-1$ \cite{Li-Ma}. The second one is to expand $\omega$ in terms of an appropriated local basis \cite{Daly2003,Huterer2005,Alam2006,Hojjati2010}. Consequently, we are 
interested in proposing a divergence-free reconstruction for $\omega$ and via a polynomial expansion, it will show new features. For example, we can expand $\omega$, 
in function of the Chebyshev polynomials $T_{n}$, with $n\in N$. These polynomials are considered as a complete orthonormal basis on the finite interval 
$[-1,1]$, and belong to the Hilbert space $L^{2}$ of real values \cite{Olivier2012}. Likewise, they have the property to be minimal approximately polynomials 
\cite{Simon2005,Martinez2008}, and possess a better advantage in terms of stability.\\
If we compare the theoretical predictions with the observational measurements, we will show the different effects of including the numerical reconstructions of $\bar{Q}$ 
and $\omega$ on the energy densities (${\bar{\Omega}}_{DE}$ and ${\bar{\Omega}}_{DM}$), on the evolution of the linear growth rate of $DM$ density perturbation 
($f\sigma_{8}$), on the effective Hubble friction term ($H_{eff}$) and on the effective Newton constant ($G_{eff}$), respectively. This will be the main aim of the 
present work.\\
On the other hand, RSD data represent a compilation of measurements of the quantity $f(z)\sigma_{8}(z)$ at different redshifts, which were obtained in a model independent 
way. These data are apparent anisotropies (effects) of the galaxy distribution (in redshift space), due to the differences of the estimates between the redshifts observed 
distances and true distances, and caused by the component along the line of sight (LOS) of the peculiar velocity of each of the galaxies (recessional velocity) 
\cite{Kaiser1987,Sargent1977,Hamilton1997}. Therefore, RSD data will provide tight constraints on the parameter space of the cosmic scenarios, and the necessary 
information to discriminate among all them \cite{Tegmark2006,Guzzo2008,Blake2011,Peacok2001,Hawkins2003,Reid2012,DeFelice2011,Appleby2012,DeFelice2012,Nishimichi2013}.\\
Furthermore, another interesting observable considered here are the measurements of the Alcock-Paczynski (AP) effect \cite{Alcock1979}. This AP test describes a distortion 
along the observed tangential and radial dimensions of objects, which are assumed as isotropic \cite{Alcock1979,Matsubara1996,Ballinger1996}. This signal depends on the 
value of the $F_{AP}$ parameter, and will be very useful to constrain cosmological models. In this work, the linear growth rate $f\sigma_{8}$ is constrained by measuring 
the RSD signal, while the dilation scale ($D_{v}$) \cite{Eisenstein2005} and $F_{AP}$ parameter \cite{Alcock1979} evaluated at an effective redshift $z_{eff}$ 
are constrained by measuring the BAO and AP signals, respectively.\\
Now then, two distinct coupled $DE$ models such as XCPL and DR are studied here, and from which we have found a determined number of different effects such as a 
reduction or enhancement on the amplitudes of ${\bar{\Omega}}_{DE}$, ${\bar{\Omega}}_{DM}$, $f\sigma_{8}$, $H_{eff}$ and $G_{eff}$ at large and small redshifts with respect 
to those found in the uncoupled models. However, these modifications should be small in order to do not have a significant impact on the matter density perturbations. 
Furthermore, important features of the universe can be obtained from these changes, and therefore these variations depend of the chosen parameterizations for $\bar{Q}$ and 
$\omega$, respectively. In this article, all our models are constrained using an analysis combined of JLA (SNe Ia) \cite{Conley2011,Jonsson2010,Betoule2014}, the growth rate 
of structure formation obtained from RSD data 
\cite{Jackson1972,Kaiser1987,Mehrabi2015,Alcock1979,Seo2008,Battye2015,Samushia2014,Hudson2013,Beutler2012,Feix2015,Percival2004,
Song2009,Tegmark2006,Guzzo2008,Samushia2012,Blake2011,Tojeiro2012,Reid2012,delaTorre2013,Planck2015}, BAO data \cite{Blake2011,Planck2015,Hinshaw2013,Beutler2011,Ross2015,Percival2010,
Kazin2010,Padmanabhan2012,Chuang2013a,Chuang2013b,Anderson2014a,Kazin2014,Debulac2015,FontRibera2014,Eisenstein1998,Eisenstein2005,Hemantha2014}, 
CMB data \cite{Planck2015,Hu-Sugiyama1996,Bond-Tegmark1997,Neveu2016} and the H data set \cite{Sharov2015,Zhang2014,Simon2005,Moresco2012,Gastanaga2009,Oka2014,Blake2012,Stern2010,Moresco2015,Busca2013}.\\
Finally, we organize this paper as follows: The background equation of motions for the energy densities are presented in section II. In section III we describe the 
reconstruction schemes for $\bar{Q}$ and $\omega$, respectively. In section IV we show the theoretical $DE$ models. In section V we studied the conditions for the crossing
of ${{\rm{I}}}_{\rm{Q}}=0$ line, and define the redshift crossing points. The perturbed equation of motions and the equations of structure formations are described in 
section VI. The current observational data and the priors considered are presented in section VII. We discuss our results in section VIII. In section IX, we conclude our 
main results.
\section{Background equations of motion}\label{Background}
In a flat Friedmann-Robertson-Walker (FRW) universe its background dynamics is described by the following set of equations for their energy densities 
(detailed calculations are found in \cite{Cueva-Nucamendi2012}, so we do not discuss them here.)
\begin{eqnarray}
\label{EoFB}{\dot{\bar{\rho}}}_{b}+3{\bar{H}}{\bar{\rho}}_{b}&=&0\;,\\
\label{EoFr}{\dot{\bar{\rho}}}_{r}+4{\bar{H}}{\bar{\rho}}_{r}&=&0\;,\\
\label{EoFDM}{\dot{\bar{\rho}}}_{DM}+3\bar{H}{\bar{\rho}}_{DM}&=&+\bar{Q}\;,\\
\label{EoFDE}{\dot{\bar{\rho}}}_{DE}+3 \left(1+{\omega}\right)\bar{H}{\bar{\rho}}_{DE}&=&-\bar{Q}\;,
\end{eqnarray}
where ${\bar{\rho}}_{b}$, ${\bar{\rho}}_{r}$, ${\bar{\rho}}_{DM}$ and ${\bar{\rho}}_{DE}$ are the energy densities of the baryon, radiation, $DM$ and $DE$, respectively. Now, 
defined the Hubble expansion rate as $\bar{H} \equiv \dot{a}/a$, and ``$\cdot$'' indicates differentiation with respect to the cosmic time $t$.\\
In what follows, we shall assume that there is not energy transfer from $DE$ ($DM$) to baryon or radiation, and among them only exist a gravitational coupling
\cite{Koyama2009-Brax2010}. For convenience, we defined the critical density $\rho_{c}\equiv 3{\bar{H}}^2/8\pi G$ and the critical density today $\rho_{c,0}\equiv 3H^2_{0}/8\pi G$ (in where
$H_{0}$ is the current value of the Hubble parameter). Considering that $A=b, r, DM, DE$, and then the normalized densities become
\begin{equation}\label{energydensity} 
{\bar{\Omega}}_{A}\equiv\frac{\bar{\rho}_{A}}{\rho_{c}}=\frac{\bar{\rho}_{A}/\rho_{c,0}}{\rho_{c}/\rho_{c,0}}=\frac{\Omega^{\star}_{A}}{E^{2}}\;,\qquad{\Omega}_{A,0}\equiv
\frac{{\rho}_{A,0}}{\rho_{c,0}}\;.
\end{equation}
The first Friedmann equation is given by
\begin{eqnarray}\label{hubble} 
E^{2}\equiv\frac{{\bar{H}}^{2}}{{H}^{2}_{0}}&=&\frac{8\pi G}{3{H}^{2}_{0}}\left({\bar{\rho}}_{b}+{\bar{\rho}}_{r}+{\bar{\rho}}_{DM}+{\bar{\rho}}_{DE}\right)\;,\nonumber\\
&&=\left[{\Omega}^{\star}_{b}+{\Omega}^{\star}_{r}+{\Omega}^{\star}_{DM}+{\Omega}^{\star}_{DE}\right]\;,
\end{eqnarray}
and with the following relation for all time
\begin{equation}\label{OmegaDE_present} 
{\bar{\Omega}}_{b}+{\bar{\Omega}}_{r}+{\bar{\Omega}}_{DM}+{\bar{\Omega}}_{DE}=1\;\;.  
\end{equation}
The scale factor $a$ is related to the redshift through $a=1/(1+z)$, from which find $dt/dz=-1/(1+z){\bar{H}}(z)$. Substituting this last relation into 
Eqs.\;(\ref{EoFB})-(\ref{EoFDE}), and solving Eqs.\;(\ref{EoFB}) and (\ref{EoFr}), we find
\begin{eqnarray}
\label{Omegabs} \Omega^{\star}_{b}(z)&=&\Omega_{b,0}{(1+z)}^{3}\;,\\
\label{Omegars} \Omega^{\star}_{r}(z)&=&\Omega_{r,0}{(1+z)}^{4}\;,\\
\label{EDFDMOmega} 
\frac{\mathrm{d}\Omega^{\star}_{DM}}{\mathrm{d}z}-\frac{3\,\Omega^{\star}_{DM}}{1+z}&=&\frac{-\Omega^{\star}_{DM}{\rm I}_{\rm Q}(z)}{1+z}\,,\quad\\
\label{EDFDEOmega} 
\frac{\mathrm{d}\Omega^{\star}_{DE}}{\mathrm{d}z}-\frac{3(1+\omega)\,\Omega^{\star}_{DE}}{1+z}&=&\frac{+\Omega^{\star}_{DM}{\rm I}_{\rm Q}(z)}{1+z}\,.
\end{eqnarray}
These equations are fundamental to determine the results within our models.
\section{Parameterizations of $\bar{Q}$ and $w$} \label{General Reconstruction}
The Chebyshev polynomials form a complete set of orthonormal functions on the interval $[-1, 1]$, and have the property to be the minimal approximating polynomials. 
It is to say, they have the smallest maximum deviation from the true function at any given order \cite{Simon2005, Cueva-Nucamendi2012}.\\
Because of the unknown of the origin and nature of the dark fluids, it is not possible to derive $\bar{Q}$ from fundamental principles, but we have the freedom of choosing 
any possible form of $\bar{Q}$ that satisfies Eqs. (\ref{EDFDMOmega}) and (\ref{EDFDEOmega}) simultaneously. Hence, we propose a phenomenological form for a varying $Q$, 
which could be definitely a function of ${\bar{\rho}}_{DM}$ multiplied both by a quantity with units of inverse of time (for instance $\bar{H}$) and by the coupling term, 
${\bar{\rm I}}_{\rm Q}$. Since ${\bar{\rm I}}_{\rm Q}$ can be modelled as a varying function of $z$ and used to measure the strength of the interaction, it can be reconstructed 
conveniently in terms of Chebyshev polynomials. Accordingly, we can look new physical properties in spite of the fact that $\bar{Q}$ may be determined by the universal 
expansion rate $\bar{H}$. Therefore, this scenario will simplify the analytic results, if we reconstruct ${\bar{\rm I}}_{\rm Q}$ from observational data. An energy exchange 
in the dark sector can be chosen as 
\begin{equation}\label{Interaction} 
{\bar{Q}}\equiv \bar{H}\bar{\rho}_{DM}\bar{{\rm I}}_{\rm Q}\;.
\end{equation}
Here, the strength of the coupling is characterized by 
\begin{equation}\label{CouplingIq} 
{\bar{\rm I}}_{\rm Q}\equiv \sum_{n=0}^{2}{{\lambda}}_{n}T_{n}\;,
\end{equation}
where the coefficients of the polynomial expansion ${\lambda}_{n}$ are free 
dimensionless parameters \cite{Cueva-Nucamendi2012}, and
\begin{equation}\label{Chebyshev1} 
T_{0}(z)=1\;,\hspace{0.3cm}T_{1}(z)=z\;,\hspace{0.3cm} T_{2}(z)=(2z^{2}-1)\;, 
\end{equation}
represent the first three Chebyshev polynomials.\\
Within the CPL model, the past evolution history may be successfully described by its EoS parameter, $\omega$, but the future evolution may not be explained, 
because $\omega$ grows increasingly, and then, encounters a divergence when $z\rightarrow -1$. That is not a physical feature. Consequently, to avoid such divergence problem
we propose here a complete phenomenological reconstruction of a smoothly varying EoS parameter, $\omega$, which can also be expanded in terms of an expansion of the 
Chebyshev polynomials such that 
\begin{equation}\label{CouplingIw} 
\omega(z)\equiv \omega_{2}+2\sum^{2}_{m=0}\frac{\omega_{m}T_{m}}{2+z^{2}}\,,
\end{equation}
where $\omega_{0}, \omega_{1}$ and $\omega_{2}$ are free dimensionless parameters. The polynomial $(1-2z^{2})^{-1}$ and the parameter $\omega_{2}$ 
were included conveniently to simplify the calculations. However, this suitable generalization should be compatible with recent observational data. Likewise, ${\omega}(z)$ 
behaves nearly linear at low redshift ${\omega}(z=0)=\omega_{0}$ and $d{\omega}/dz|_{z=0}=\omega_{1}$, whereas in the high redshift regime ${\omega}(z) \simeq 5{\omega_{2}}$.\\
The Chebyshev polynomials of order $m=2$ were defined by Eq. (\ref{Chebyshev1}). Thereafter, using numerical simulations we will compute the best fitted values for 
${\lambda}_{0}$, ${\lambda}_{1}$, ${\lambda}_{2}$, $\omega_{0}$, $\omega_{1}$ and $\omega_{2}$, respectively.
\section{Dark\,energy\,models} \label{DE_models}
 \subsection{$\Lambda$CDM model}\label{LCDM}
In this scenario, the function $E^{2}$ is defined fixing both $\omega(z)=-1$, and $\bar{Q}(z)=0$ into Eqs. (\ref{Omegabs})-(\ref{EDFDEOmega})
\begin{equation}\label{hubble_LCDM}
E^{2}=\biggl[\Omega^{\star}_{b}(z)+\Omega^{\star}_{r}(z)+\Omega_{DM,0}{(1+z)}^{3}+\Omega_{DE,0}\biggr]\;.
\end{equation}
\subsection{CPL model}\label{CPL}
Within this model, $E^{2}$ is determined replacing both $\omega(z)=\omega_{0}+\omega_{1}[z/(1+z)]$, where $\omega_{0}$, $\omega_{1}$ are real parameters, 
and $\bar{Q}(z)=0$ into Eqs. (\ref{Omegabs})-(\ref{EDFDEOmega})
\begin{eqnarray}\label{hubble_CPL}E^{2}&=&\Biggl[\Omega_{b,0}{(1+z)}^{3}+\Omega_{r,0}{(1+z)}^{4}+\Omega_{DM,0}{(1+z)}^{3} \nonumber\\
&& +\Omega_{DE,0}{(1+z)}^{3(1+\omega_{0}+\omega_{1})}{{\rm exp}}\biggl({\frac{-3\omega_{1}z}{1+z}}\biggr)\Biggr]\;.\end{eqnarray}
\subsection{XCPL model}\label{XCPL}
Firstly a coupled model can be defined putting both $\omega=\omega_{0}+\omega_{1}(z/{1+z})$, where $\omega_{0}$, $\omega_{1}$ are real free 
parameters, and ${\bar{Q}}(z)$ given by Eqs. (\ref{Interaction}) and (\ref{CouplingIq}) into Eqs. (\ref{Omegabs})-(\ref{EDFDEOmega}).
The explicit form for $\Omega^{\star}_{DM}$ and $\Omega^{\star}_{DE}$ are reached by solving Eqs. (\ref{EDFDMOmega}) and (\ref{EDFDEOmega}),
respectively.
\begin{eqnarray}
\label{DM_XCPL}\Omega^{\star}_{DM}(z)=(1+z)^{3}{\Omega_{DM,0}}{\rm exp}\biggl[{\frac{-z_{max}}{2}\sum_{n=0}^{2}\lambda_{n}I_{n}(z)}\biggr],\qquad\\
\label{DE_XCPL}\Omega^{\star}_{DE}(z)=(1+z)^{3(1+{\omega}_{0}+{\omega}_{1})}\biggl[{\Omega_{DE,0}}{\rm exp}\left(\frac{-3\omega_{1}z}{1+z}\right)+\;\quad\qquad\nonumber\\
\frac{z_{max}}{2}\Omega_{DM,0}{\rm exp}\left(\frac{3\omega_{1}}{1+z}\right)\sum_{n=0}^{2}\lambda_{n}S_{n}(z,\bar{\omega})\biggr]\;.\quad\qquad
\end{eqnarray}
The following average integrals are also defined
\begin{eqnarray}
\label{Average_In}\int_{0}^{z}\frac{T_{n}(\tilde{x})}{(1+\tilde{x})}d\tilde{x} & \approx & \frac{z_{max}}{2}I_{n}(z)\;,\\
\label{Average_Sn}\int_{0}^{z}\frac{T_{n}(\tilde{x})A(\tilde{x})B(\tilde{x})}{(1+\tilde{x})^{(1+3{\omega}_{0}+3{\omega}_{1})}}d\tilde{x} &\approx& 
\frac{z_{max}}{2}S_{n}(z,\bar{\omega})\;,
\end{eqnarray}
where we also defined the following expressions for all $n\in[0,2]$ (see Appendix \ref{integralIn} and \cite{Cueva-Nucamendi2012})
\begin{eqnarray}
A(\tilde{x})&=&{{\rm exp}}\left(\frac{-z_{max}}{2}{\sum_{n=0}^{2}}\lambda_{n}I_{n}(\tilde{x})\right)\;,\nonumber\\
\label{a1}\tilde{A}(\tilde{x})&=&{{\rm exp}}\left(\frac{-z_{max}}{2}{\sum_{n=0}^{2}}\lambda_{n}\tilde{I}_{n}(\tilde{x})\right)\;,\nonumber\\
B(\tilde{x})&=&{{\rm exp}}\left(\frac{-3\omega_{1}}{1+\tilde{x}}\right),\quad\tilde{B}(\tilde{x})={{\rm exp}}\left(\frac{-3\omega_{1}}{a+b\tilde{x}}\right),\nonumber\\
I_{n}(z)&\equiv&\int_{-1}^{x} \frac{T_{n}(\tilde{x})}{(a+b\tilde{x})}d\tilde{x}\;,\nonumber\\
S_{n}(z,\bar{\omega})&\equiv&\int_{-1}^{x}\frac{T_{n}(\tilde{x})\tilde{A}(\tilde{x})\tilde{B}(\tilde{x})}{(a+b\tilde{x})^{(1+3\omega_{0}+3\omega_{1})}}d\tilde{x}\;,
\end{eqnarray}
with the quantities
\begin{equation*}
x\equiv 2(z/z_{max})-1\;,\quad a\equiv1+\frac{z_{max}}{2}\;,\quad b\equiv\frac{z_{max}}{2}\;,
\end{equation*}
where $z_{max}$ is the maximum value for $z$, and in which the observations are possible such that $\tilde{x}\in [-1, 1]$ and
$\vert T_{n}(\tilde{x})\vert\leq 1$, respectively.\\
Therefore, the function $E^{2}$ can be constructed from Eqs. (\ref{hubble}), (\ref{Omegabs}), (\ref{Omegars}), (\ref{DM_XCPL}) and (\ref{DE_XCPL}).
\subsection{DR model}\label{DR}
Secondly another coupled model can be modeled setting 
\begin{equation}\label{DE_DR}
\omega(z)\equiv \omega_{2}+2\sum^{2}_{m=0}\frac{\omega_{m}T_{m}}{2+{z}^{2}}\,, 
\end{equation}
where $\omega_{0}$, $\omega_{1}$ and $\omega_{2}$ are real parameters. Moreover, ${Q}(z)$ is given by Eqs. (\ref{Interaction}) and (\ref{CouplingIq}).
The analytic form for $\Omega^{\star}_{DM}$ can be reached by solving Eq. (\ref{EDFDMOmega}). For this model Eq. (\ref{DM_XCPL}) represents the solution of
Eq. (\ref{EDFDMOmega}). Instead, the solution of Eq. (\ref{EDFDEOmega}) can be obtained by numerical integration, from using Eq. (\ref{a1}) and Appendix \ref{integralIn},
\begin{equation}\label{DE_DR}
\Omega^{\star}_{DE}(z)=C(z)+D(z)\int_{-1}^{x}{\sum_{n=0}^{2}\lambda_{n}T_{n}(\tilde{x})}\frac{F(\tilde{x})}{G(\tilde{x})}d\tilde{x}\;,
\end{equation}
where the following relations are defined
\begin{widetext}
\begin{eqnarray}
C(z)&=&C_{0}(z).C_{1}(z).C_{2}(z)\;,\;\;\;D(z)=D_{0}(z).D_{1}(z).D_{2}(z)\;,\;\;\;F(\tilde{x})=F_{0}(\tilde{x}).F_{1}(\tilde{x}).F_{2}(\tilde{x})\,,\nonumber\\
C_0(z)&=&\Omega_{DE,0}(1+z)^{3(1+\omega_{2})+A_0}\;,\;\;\;C_1(z)=\biggl[\frac{(2z-z_{max})^{2}+2(z_{max}+2)^{2}}{3z_{max}^{2}+8(1+z_{max})}\biggr]^{A_1}\;,\nonumber\\
C_2(z)&=&{\rm exp}\Biggl[A_{2}\sqrt{2}\left(\arctan\left[\frac{\sqrt{2}[2z-z_{max}]}{2z_{max}+4}\right]+\arctan\left[\frac{\sqrt2z_{max}}{2z_{max}+4}\right]\right)\Biggr]\;,\nonumber\\
D_0(z)&=&0.5z_{max}\Omega_{DM,0}(1+z)^{2\omega_{0}-2\omega_{1}+5\omega_{2}+3}(2+z^{2})^{\omega_{1}-\omega_{0}+5\omega_{2}}\;,\nonumber\\
D_1(z)&=&{\rm exp}\biggl[{\sqrt{2}}\biggl(\omega_{0}+2\omega_{1}-5\omega_{2}\biggr)\arctan\left(0.5{\sqrt{2}}z\right)\biggr]\;,\;\;\;D_2(z)={\rm exp}\biggl[\arctan\left(\frac{\sqrt{2}z_{max}}{4+2z_{max}}\right)J_{2}\biggr]\;,\nonumber\\
F_0(\tilde{x})&=&\frac{{\rm exp}\biggl[-\sqrt{2}\biggl(\omega_{0}+2\omega_{1}-5\omega_{2}\biggr)\arctan\biggl[0.25\sqrt{2}z_{max}(1+\tilde{x})\biggr]\biggr]}{\biggl[\left(\tilde{x}^{2}+2\right)(0.5z_{max})^{2}+2\left(1+z_{max}\right)\biggr]^{\omega_{1}-\omega_{0}+5\omega_{2}}}\;,\nonumber\\
F_1(\tilde{x})&=&\biggl[\frac{\tilde{x}^{2}+2+8[z_{max}^{-2}+z_{max}^{-1}]}{3+8[z_{max}^{-2}+z_{max}^{-1}]}\biggr]^{J_{1}}\;,\;\;\;F_2(\tilde{x})={\rm exp}\biggl[\arctan\left(\frac{\sqrt2z_{max}\tilde{x}}{2z_{max}+4}\right)J_{2}\biggr]\;,\nonumber\\
G(\tilde{x})&=&\biggl[1+0.5z_{max}(1+\tilde{x})\biggr]^{2(\omega_{0}-\omega_{1}+\omega_{2})-J_{0}}\;.
\end{eqnarray}
\end{widetext}
Within this model, the function $E^{2}$ can be constructed from Eqs. (\ref{hubble}), (\ref{Omegabs}), (\ref{Omegars}), (\ref{DM_XCPL}) and (\ref{DE_DR}). 
The basic analytical expressions for $A_n(x)$ and $J_{n}(\tilde{x})$ (when $n=0,1,2$) are shown in Appendix \ref{integralIn}.
\section{Crossing of ${\bar{\rm I}}_{\rm Q}=0$ line with a minimal derivative coupling.}\label{crossingIq}
Let us now proceed with the calculation of the redshift crossing points, and analyze the behavior of ${\bar{\rm I}}_{\rm Q}$ and its derived. 
From Eqs. (\ref{Interaction}), (\ref{CouplingIq}) and (\ref{Chebyshev1}), we note that exist real values of $z$ that lead to ${\bar{\rm I}}_{\rm Q}(z_{crossing})=0$
\begin{equation}
\label{crossIq} {\bar{\rm I}}_{\rm Q}(z_{crossing})=2\lambda_{2}{z^{2}_{crossing}}+\lambda_{1}{z_{crossing}}+(\lambda_{0}-\lambda_{2})=0.
\end{equation}
in where the $z_{crossing}$ denotes the redshift crossing points of ${\bar{\rm I}}_{\rm Q}=0$ line.\\
Then, the solution to Eq. (\ref{crossIq}) is given by
\begin{equation}
\label{ZcrossIq} {z}_{crossing}=-\frac{\lambda_{1}}{4\lambda_{2}}\pm\sqrt{\left(\frac{\lambda_{1}}{4\lambda_{2}}\right)^{2}-\left(\frac{\lambda_{0}-\lambda_{2}}{2\lambda_{2}}\right)}.
\end{equation}
From here, we note that these results depend of the choice for ${\bar{\rm I}}_{\rm Q}$. However, we are interested in the case where $\bar{Q}=0$; in particular, this happens
when ${\bar{\rm I}}_{\rm Q}=0$. Furthermore, the reverse situation is also possible. In this discussion, to guarantee the possibility of the crossing of 
${\bar{\rm I}}_{\rm Q}=0$ line we must explore the function ${\mathrm{d}{\bar{\rm I}}_{\rm Q}}/{\mathrm{d}z}$. Now, we consider the possibility to have 
${\bar{\rm I}}_{\rm Q}=0$ and ${\mathrm{d}{\bar{\rm I}}_{\rm Q}}/{\mathrm{d}z}|_{z_{crossing}}$ could be zero or different of zero. Then, from Eqs. (\ref{CouplingIq}) and 
(\ref{Chebyshev1}), we obtain
\begin{equation}
\label{dcrossIqdz} \frac{\mathrm{d}{\bar{\rm I}}_{\rm Q}}{\mathrm{d}z}=\lambda_{1}+4\lambda_{2}z\,.
\end{equation}
Substituting Eq. (\ref{ZcrossIq}) into Eq. (\ref{dcrossIqdz}), we get the following
\begin{equation}
\label{dIqdznull2} \frac{\mathrm{d}{\bar{\rm I}}_{\rm Q}}{\mathrm{d}z}|_{z_{crossing}}=\pm\frac{1}{4\lambda_{2}}\sqrt{{\lambda_{1}}^{2}-8\lambda_{2}\left(\lambda_{0}-\lambda_{2}\right)}\neq 0\,,\\
\end{equation}
Now, if we consider the possibility to have ${\bar{\rm I}}_{\rm Q}(z_{crossing})=0$ and ${\mathrm{d}{\bar{\rm I}}_{\rm Q}}/{\mathrm{d}z}|_{z_{crossing}}=0$, it means that
the impossibility for having ${\bar{\rm{I}}}_{\rm{Q}}$ over ${\bar{\rm{I}}}_{\rm{Q}}=0$ line. By contrast, the only possibility for a crossing corresponds to
\begin{equation}
\label{dIqdznull}
{\bar{\rm I}}_{\rm Q}|_{z_{crossing}}=0\,\qquad \frac{\mathrm{d}{\bar{\rm I}}_{\rm Q}}{\mathrm{d}z}|_{z_{crossing}}\neq 0.
\end{equation}
From Eq. (\ref{dIqdznull2}), we impose the following restraint for the avoidance of imaginary values in ${\lambda_{1}}$
\begin{equation}
\label{dIqdznull3} 
{\lambda_{1}}\geq \sqrt{8\lambda_{2}\left(\lambda_{0}-\lambda_{2}\right)}\,\,\cup\,\,{\lambda_{1}}\leq -\sqrt{8\lambda_{2}\left(\lambda_{0}-\lambda_{2}\right)}\,,
\end{equation}
which can be rewritten as 
\begin{equation}
\label{dIqdznull4} 
\mid \lambda_{1} \mid \geq \sqrt{8\lambda_{2}\left(\lambda_{0}-\lambda_{2}\right)}\,.
\end{equation}
Similarly, from Eq. (\ref{dIqdznull4}) to guarantee real values with physical sense, we impose
\begin{eqnarray}
\label{dIqdznull5} 
8\lambda_{2}\left(\lambda_{0}-\lambda_{2}\right)\geq 0\,\rightarrow\,\\
\label{dIqdznull6} 
\left(\lambda_{2}\geq 0\right)\,\cap\,\left(\lambda_{0}\geq\lambda_{2}\right)\,\cup\\
\label{dIqdznull7}
\left(\lambda_{2}\leq 0\right)\,\cap\,\left(\lambda_{0}\leq\lambda_{2}\right)\,.
\end{eqnarray}
The values of $z_{crossing}$ for the coupled $DE$ models are given in Table \ref{crossing_state}. Let us make some commentaries about the signs of $\bar{Q}$, 
${\bar{\rm I}}_{\rm Q}$ and $\mathrm{d}{\bar{\rm I}}_{\rm Q}/\mathrm{d}z$, respectively. In general, if the parameters $\lambda_{0}$, $\lambda_{1}$ and $\lambda_{2}$ hold 
positive or negative values, then ${\bar{\rm I}}_{\rm Q}$ and $\mathrm{d}{\bar{\rm I}}_{\rm Q}/\mathrm{d}z$ will be ambiguously positive, negative or zero, in any epoch of 
the universe. 
From Eqs. (\ref{dcrossIqdz}), (\ref{dIqdznull4}), (\ref{dIqdznull6}) and (\ref{dIqdznull7}), if $\lambda_{0}$ and $\lambda_{2}$ are both positive or are both negative, 
then $\mathrm{d}{\bar{\rm I}}_{\rm Q}/\mathrm{d}z$ could be positive or negative. Moreover, $\mathrm{d}{\bar{\rm I}}_{\rm Q}/\mathrm{d}z$ may be zero when $\lambda_{0}$, 
$\lambda_{1}$, and $\lambda_{2}$ are all zero (i.e. uncoupled $DE$ models) or when $\mid \lambda_{1} \mid=\sqrt{8\lambda_{2}\left(\lambda_{0}-\lambda_{2}\right)}$. Here we 
can describe the signs of $\bar{Q}$ and ${\bar{\rm I}}_{\rm Q}$, choosing positive values for $\lambda_{0}$ either with negative values of $\lambda_{1}$ and positive 
values of $\lambda_{2}$ or with positive values of $\lambda_{1}$ and negative (positive) values of $\lambda_{2}$, in determined redshift ranges.\\
\section{Perturbed equations of motion.}\label{structure}
\subsection{Perturbed equations of motion for coupled DE models.}\label{perturbedEq}
Let us consider a spatially flat universe with scalar perturbations about the background. In the absence of the anisotropic stress, the perturbed line element in the Newtonian 
gauge is given by \cite{cabral2009}
\begin{equation}
\label{metricFRW} ds^{2}=-\left(1+2\phi\right)dt^{2}+a^{2}(t)\left(1-2\phi\right)d{\vec{r}}^{\,\,2}\,,
\end{equation}
where $\phi$ is the gravitational potential, $a(t)$ is the scale factor, $U^{\mu}=\delta^{\mu}_{0}$ is the background four velocity, $U^{\mu}_{A}=(1-\phi,\partial^{k}v_{A})$ 
or $U_{\mu}^{A}=(-(1+\phi),\partial_{k}v_{A})$ is the $A$ perfect-fluid four velocit, and $v_{A}$ is the $A$ fluid peculiar velocity potential. In addition, to avoid a 
momentum flux relative to $U^{\mu}_{A}$, we define the total four velocity $U^{\mu}$ as \cite{cabral2009}
\begin{equation}\label{fourvelocity}
U^{\mu}=(1-\phi,\partial^{k}v)\,,
\end{equation}
where the total velocity potential $v$ is given by \cite{cabral2009}
\begin{equation}\label{velocitypotential}
(p+\rho)v = \sum_{A}(\bar{\rho}_{A}+\bar{p}_{A})v_{A},
\end{equation}
with $\rho=\sum_{A}{\bar\rho}_{A}$, $P=\sum_{A}{\bar{P}}_{A}$ and $A=DM, DE, b, r$. This is the choice of $v$ that we will use from now on.\\
Thus, the $A$ fluid energy-momentum tensor is \cite{cabral2009}
\begin{equation}\label{A_Energy_momentum_tensor}
{T_{A}}^{\mu\nu}=({\bar{\rho}}_{A}+{\bar{P}}_{A}){U^{\mu}_{A}}{U^{\nu}_{A}}+{g}^{\mu\nu}{\bar{P}}_{A}\,,
\end{equation}
where $\rho_{A}=\bar{\rho}_{A}+\delta\rho_{A}$ and $P_{A}=\bar{P}_{A}+\delta P_{A}$.
The covariant form of energy-momentum transfer is satisfied for the whole system, while for each component we have \cite{cabral2009}
\begin{equation}\label{Energy_momentum_transfer}
\nabla_{\nu}{T_{A}}^{\mu\nu}=F_{A}^{\mu}\,,\qquad \sum_{A}F_{A}^{\mu}=0\,,
\end{equation}
where $F_{A}^{\mu}$ describe the interaction, $F_{A}^{\mu}=0$ for $A=(b,\,r)$ in the late universe, and $F_{DM}^{\mu}=-F_{DE}^{\mu}\neq0$.
A general $F_{A}^{\mu}$ relative to the total four velocity can be split as \cite{cabral2009}
\begin{eqnarray}
 \label{decomposeF} F_{A}^{\mu}=Q_{A}U^{\mu}+ N^{\mu}_{A}\,,\\
 \label{decomposeQ} Q_{A}=\bar{Q}_{A}+\delta Q_{A}\,,\\
 \label{NA} u_{\mu}N^{\mu}_{A}=0\,,
\end{eqnarray}
where $Q_{A}$ is the energy density transfer and $N^{\mu}_{A}$ is the momentum density transfer rate, relative to $U^{\mu}_{A}$. Here, we also choose 
$N^{\mu}_{A}=(0,\partial^{k}f_{A})$, where $f_{A}$ is a momentum transfer potential. Then, from Eq. (\ref{decomposeF}) we find that \cite{cabral2009}
\begin{eqnarray}
\label{componentFo} F^{A}_{0}&=&-\left[\bar{Q}_{A}(1+\phi)+\delta Q_{A}\right]\,,\\
\label{componentFk} F^{A}_{k}&=&a^{2}\partial_{k}\left(f_{A}+\bar{Q}_{A}v\right).  
\end{eqnarray}
The perturbed energy transfer $F^{A}_{0}$ includes a metric perturbation term ${\bar{Q}}_{A}\phi$ and a perturbation $\delta Q_{A}$. In addition, we stress that 
the perturbed momentum transfer $F^{A}_{k}$ is made up of two parts: the momentum transfer potential ${\bar{Q}}_{A}v$ that arises from energy transport along the total velocity 
and the intrinsic momentum transfer potential $f_{A}$. Hence, the total energy-momentum conservation implies that
\begin{equation}\label{conservation}
 \sum_{A}Q_{A} = \sum_{A}\delta Q_{A}= \sum_{A}f_{A}=0\,.
\end{equation}
\subsection{Structure formation in coupled $DE$ models.}\label{formation}
The general evolution equations for the dimensionless density perturbation $\delta_{A}=\delta{\rho}_{A}/\bar{\rho}_{A}$ is given by
\cite{Malik2003,cabral2009,cabral2010}
\begin{widetext}
\begin{eqnarray}\label{contrast1}
\dot{\delta}_{A}+3\bar{H}c^{2}_{SA}\delta_{A}-(1+{w}_{A})\frac{k^2}{a}v_{A}-3\bar{H}
\left[3\bar{H}(1+{w}_{A})(c^{2}_{SA}-{w}_{A})+ \dot{{w}}_{A}\right]v_{A} 
-3(1+{{w}}_{A}){\dot \phi}&=&\frac{{\delta Q}_{A}}{{\bar{\rho}}_{A}}+\nonumber\\
\frac{{\bar{Q}}_{A}}{{\bar{\rho}}_{A}}\left[\phi-\delta_{A}-3a{\bar{H}}(c^{2}_{SA}-{w}_{A})v_{A}\right] ,
\end{eqnarray}
\end{widetext}
also the velocity perturbation equation takes the form
\begin{widetext}
\begin{eqnarray}\label{euler1} 
{{\dot{v}}_{A}}+{\bar{H}}(1-3{c^{2}_{SA}}){{v}_{A}}+\frac{c^{2}_{SA}}{a(1+{{{w}}_{A}})}{{\delta}_{A}}
+\frac{\phi}{a}&=&\frac{1}{(1+{{{w}}_{A}}){{\bar{\rho}}_{A}}}\left[{{\bar{Q}}_{A}}\left(v-(1+{c^{2}_{SA}}){{v}_{A}}\right)+{f_{A}}\right]\,,
\end{eqnarray}
\end{widetext}
and the relativistic Poisson equation is
\begin{equation}\label{poisson1} 
\frac{k^{2}\phi}{a^{2}}=-3{\bar{H}}{\dot{\phi}}-3{{\bar{H}}^{2}}\phi-{4\pi}G\left({{\bar{\rho}}_{A}}{{\delta}_{A}}\right)\,. 
\end{equation}
where $G$ is Newton's constant.\\
We now consider that $DE$ does not cluster on sub-Hubble scales ${{H}}\ll{k/a}$, and therefore, we could ignore $\dot{\delta}_{DE}$ from Eq. (\ref{contrast1}).
Moreover, to avoid the nonphysical sound speed, we choose $c_{DE}^{2}=1$ \cite{valiviita2008,Majerotto2010,Clemson2012}.\\
Similarly, we also assume that the dynamical effects of the gravitational potential $\phi$, its time derivative $\dot{\phi}$ and the transfer of energy between 
baryons and radiation, may be neglected relative to $DM$ perturbation, $\delta_{DM}$. Here, we also consider the case where $DM$ component behaves as dust with an EoS 
parameter $\omega_{DM}=0$ and with a $DM$ sound speed $c^{2}_{DM}=0$.\\
In the linear approach Eqs. (\ref{contrast1}) and (\ref{euler1}) describe the evolution of the $DM$ perturbation $(\delta_{DM}\ll1)$, which can be rewritten as
\cite{Malik2003,cabral2009,Koyama2009-Brax2010}
\begin{equation}\label{ContinuityDM} 
{\dot \delta}_{DM}-\frac{k^{2}}{a}v_{DM}=\frac{{\delta Q}_{DM}}{\bar{\rho}_{DM}}-\frac{{{\bar Q}_{DM}}\delta_{DM}}{\bar{\rho}_{DM}}\,,
\end{equation} 
\begin{equation}\label{EulerDM} 
{{\dot v}_{DM}}+\bar{H}{v_{DM}}+\frac{\phi}{a}=\frac{1}{{\bar{\rho}}_{DM}}\left[{\bar{Q}}_{DM}(v-{{v}_{DM}})+{f}_{DM}\right]\,.
\end{equation} 
In this linear regime the Poisson equation reduces to
\begin{equation}\label{PoissonDM}
\frac{k^{2}\phi}{a^{2}}= -4\pi G\left(\bar{\rho}_{DM}\delta_{DM}+\bar{\rho}_{b}\delta_{b}\right) . 
\end{equation}
In order to satisfy the weak equivalence principle and ensure that the particles of the $DM$ can follow geodesics, we need to impose the following condition
\begin{equation}\label{f_dm} 
{f_{DM}}=-{{\bar{Q}}_{DM}}(v-{v_{DM}})\,.
\end{equation}
Here, Eq. (\ref{euler1}) for $DM$ component yields
\begin{equation}\label{EulerDMFinal} 
{{\dot v}_{DM}}+\bar{H}{v_{DM}}+\frac{\phi}{a}=0\,.
\end{equation} 
This expression means that the $DM$ velocity perturbation is not affected by the interaction with $DE$. Then, we provide a phenomenological covariant choice of the energy 
exchange four-vector
\begin{equation}\label{FDM}
F_{DM}^{\mu}=-F_{DE}^{\mu}=Q_{DM}U_{DM}^{\mu}=(\bar{Q}_{DM}+\delta Q_{DM})U_{DM}^{\mu}\,,
\end{equation}
in where one takes
\begin{eqnarray}
\label{parts1_Exchange_energy} {{\bar{Q}}_{DM}}&=&-\bar{H}{\bar{\rho}}_{DM}{\bar{\rm I}}_{\rm Q}\,,\\
\label{parts2_Exchange_energy}{\delta Q}_{DM}&=&-{\bar\rho}_{DM}{\bar{\rm I}}_{\rm Q}\delta{H}-
\bar{H}{\bar{\rm I}}_{\rm Q}{{\bar{\rho}}_{DM}}{\delta}_{DM}-\bar{H}{\bar\rho}_{DM}{\delta{\rm I}}_{\rm Q}.\qquad
\end{eqnarray}
We impose the following conditions ${{\delta{\rm I}}_{\rm Q}}\ll{\delta_{DM}}$ and ${{\delta{\rm I}}_{\rm Q}}\ll{\delta H}$ to generate $DM$ cosmic structure formation 
(In a forthcoming article we will extend our study, by considering other relations between ${{\delta{\rm I}}_{\rm Q}}$, ${\delta_{DM}}$ and $\delta{H}$.
It is beyond the scope of the present paper), so Eq. (\ref{parts2_Exchange_energy}) becomes
\begin{equation}
\label{parts_Exchange_energy}{\delta Q}_{DM}=-{\bar\rho}_{DM}{\bar{\rm I}}_{\rm Q}\delta{H}-
\bar{H}{\bar{\rm I}}_{\rm Q}{{\bar{\rho}}_{DM}}{\delta}_{DM}\,.
\end{equation}
Considering that $DM$ is more concentrated than the baryon component in the universe, also that $DE$ does not cluster on sub-Hubble scales, using Eq. (\ref{hubble}) 
and the relation $4\pi G{\bar\rho}_{DM}=1.5{\bar{H}}^{2}{{\bar{\Omega}}_{DM}}$, we have
\begin{equation}\label{parts3_Exchange_energy}
\delta H \approx \frac{\bar{H}}{2}\bar{\Omega}_{DM}\delta_{DM}\,,
\end{equation}
and then,
\begin{equation}\label{parts4_Exchange_energy}
\frac{{\delta Q}_{DM}}{\bar{\rho}_{DM}}-\frac{{\bar{Q}}_{DM}{\delta_{DM}}}{\bar{\rho}_{DM}}=\frac{\bar{H}}{2}\bar{\Omega}_{DM}{\bar{\rm I}}_{\rm Q}\delta_{DM}\,.
\end{equation}
From Eqs. (\ref{EulerDM}), (\ref{PoissonDM}), (\ref{EulerDMFinal}) and deriveting Eq. (\ref{ContinuityDM}) with respect to $t$, we find the evolution of matter density 
perturbations ${\delta}_{DM}(t)$
\begin{equation}\label{growth_rate} 
{{\ddot{\delta}}_{DM}}+(2\bar{H}-\frac{1}{2}\bar{\Omega}_{DM}\bar{H}{\bar{\rm I}}_{\rm Q}){\dot{\delta}}_{DM}-\frac{3}{2}\bar{H}^{2}\bar{\Omega}_{DM}\delta_{DM}G_{eff}=0,  
\end{equation}
When ${\bar{\rm I}}_{\rm Q}=0$, Eq. (\ref{growth_rate}) could be turner into the standard evolution of $DM$ density perturbations.
From this equation the quantity $G_{eff}$ is an effective gravitational strength (effective Newton constant), defined as
\begin{equation}\label{Geff} 
G_{eff}\equiv G\left[1+\frac{{\bar{\rm I}_{\rm Q}}^{ 2}}{3}+\frac{\dot{\bar{\rm I}}_{\rm Q}}{3H}+\frac{{\bar{\rm I}}_{\rm Q}}{6}(1+3{w}\bar{\Omega}_{DE})\right].
\end{equation}
which can be understood as a self attractive force acting on the $DM$ density perturbation and quantifies the modifications to gravity due to the effects of 
${\bar{\rm I}}_{\rm Q}$ and $\omega$ functions.\\
Here, we also define the effective Hubble friction term as
\begin{equation}\label{Heff} 
H_{eff}\equiv \bar{H}\left(1-0.25\bar{\Omega}_{DM}{\bar{\rm I}}_{\rm Q}\right).
\end{equation}
which acts as a frictional force that slows down (reduces) the growth of cosmic structure.\\
It is useful to rewrite Eq. (\ref{growth_rate}) in redshift space as
\begin{widetext}
\begin{eqnarray}\label{growth_ratez} {\delta}_{DM}^{''}+\frac{\left(1+3{\omega}{\Omega}_{DE}+\bar{\Omega}_{DM}\bar{\rm I}_{\rm Q}\right)}{2(1+z)}{\delta}_{DM}^{'}
-\frac{3\bar{\Omega}_{DM}G_{eff}}{2(1+z)^{2}}\delta_{DM}=0\\
\label{geff} G_{eff}\equiv G\left[1+\frac{{\bar{\rm I}_{\rm Q}}^{ 2}}{3}-\frac{(1+z)}{3}\frac{d\bar{{\rm I}}_{\rm Q}}{dz}+\frac{{\bar{\rm I}}_{\rm Q}}{6}(1+3{w}\bar{\Omega}_{DE})\right].
\end{eqnarray}
\end{widetext}
This equation can be solved numerically, considering that 
\begin{eqnarray}\label{conditions}
 \delta(z=0)=1\,,\\
 \delta^{'}(z=0)=-\Omega_{DM}(z=0)^{\gamma(z=0)}\,,
\end{eqnarray}
where $\gamma$ is a some unknown function of $z$ so-called the growth index of the linear matter fluctuations. 
In the linear regime, it is convenient to define the quantity
\begin{equation}\label{f} 
f\equiv\frac{dln\delta}{d{\ln a}}=-(1+z)\frac{d{\ln}\delta}{dz}\,, 
\end{equation}
called the growth factor of $DM$ density perturbations. Then, Eq. (\ref{growth_ratez}) can be rewritten in function of $f$ as 
\begin{widetext}\begin{eqnarray}\label{growth_factor} 
f^{2}+\frac{1}{2}\left(1-3{{\omega}}{\bar{\Omega}}_{DE}-{\bar{\Omega}}_{DM}{\bar{\rm I}}_{\rm Q}\right)f-(1+z)\frac{df}{dz}\equiv\frac{3}{2}\frac{G_{eff}}{G}{\bar{\Omega}}_{DM}
\end{eqnarray}
\end{widetext}
This previous equation can be solved numerically, taking into account the condition $f(0)={\Omega_{DM}(z=0)}^{\gamma(z=0)}$.\\ 
In full generality, we define the growth index of $DM$ perturbations $\gamma$ through the following ansatz
\begin{equation}\label{f}
f\equiv{\bar{\Omega}}_{DM}^{\gamma(z)}(z)\,.
\end{equation}
From Eq. (\ref{f}), we find that
\begin{equation}\label{fa}
\frac{df}{dz}={\bar{\Omega}}_{DM}^{\gamma}\biggl[\frac{d\gamma}{dz}\ln{{\bar{\Omega}}_{DM}}+\frac{\gamma}{{\bar{\Omega}}_{DM}}\frac{d{\bar{\Omega}}_{DM}}{dz}\biggr]\,.  
\end{equation}
Now, using Eqs. (\ref{EoFB})-(\ref{EoFDE}), we obtain
\begin{equation}\label{fb}
\frac{d{\bar{\Omega}}_{DM}}{dz}=-\frac{{\bar{\Omega}}_{DM}}{(1+z)}\biggl[{\bar{\rm I}}_{\rm Q}+3{\bar{\Omega}}_{DE}\omega+\Omega_{r}\biggr]\,. 
\end{equation}
Substituting Eq. (\ref{fb}) into Eq. (\ref{fa}), we get   
\begin{equation}\label{fc}\frac{df}{dz}=\frac{{{\bar{\Omega}}_{DM}}^{\gamma}}{(1+z)}\biggl[{(1+z)}\frac{d\gamma}{dz}\ln{{{\bar{\Omega}}_{DM}}}-
\gamma\left({{\bar{\rm{I}}}_{\rm Q}}+3\omega{{\bar{\Omega}}_{DE}}+{{\bar{\Omega}}_{r}}\right)\biggr]\,.
\end{equation}
Let us now to equal Eqs. (\ref{growth_factor}) and (\ref{fc}) to obtain the general evolution equation for the growth index $\gamma$
\begin{widetext}
\begin{eqnarray}\label{newgrowth_index}
\frac{d\gamma}{dz}=\frac{1}{(1+z)\ln{{{\bar{\Omega}}_{DM}}}}\biggl[{\bar{\Omega}}_{DM}^{\gamma}+\frac{1}{2}+\gamma({{\bar{\rm I}}_{\rm Q}}+{{{\bar{\Omega}}_{r}}})
-\frac{{{{\bar{\Omega}}_{DM}}}{{\bar{\rm I}}_{\rm Q}}}{2}+3\omega(1-{\bar{\Omega}}_{DM}-{\bar{\Omega}}_{b}-{\bar{\Omega}}_{r})(\gamma-\frac{1}{2})+\nonumber\\
\frac{{\bar{\Omega}}_{DM}^{1-\gamma}}{2}\left((1+z)\frac{d{{\bar{\rm I}}_{\rm Q}}}{dz}-\frac{{{\bar{\rm I}}_{\rm Q}}}{2}-{{{\bar{\rm I}}_{\rm Q}}}^{2}-3-\frac{3}{2}
\omega{{\bar{\rm I}}_{\rm Q}}\left(1-{\bar{\Omega}}_{DM}-{\bar{\Omega}}_{b}-{\bar{\Omega}}_{r}\right)\right)\biggr]\,.
\end{eqnarray}
\end{widetext}
This equation can be solved numerically by considering the condition that $\gamma(z=0)=\gamma_{0}$.\\
The parameterization given by Eq. (\ref{f}) is important to simplify rapidly the numerical calculations of Eqs. (\ref{growth_factor}) and (\ref{newgrowth_index}). 
Therefore, the $DM$ linear growth factor normalized to unity at the present epoch is given by
\begin{equation}\label{g}
g(z)=\frac{\delta_{DM}(z)}{\delta_{DM}(0)}=exp\left(-{\int^{z}_{0}}\bar{\Omega}_{DM}^{\gamma}(z)\frac{dz}{(1+z)}\right) \,,
\end{equation}
where $z$ is the redshift of the universe in which the $DM$ component dominates the universe (in this work for convenience used $z=10$).\\
Let us stress that by solving numerically Eqs. (\ref{growth_ratez}) and (\ref{growth_factor}) we can calculate ${\delta}_{DM}$ and $f$, respectively.\\ 
On the other hand, the root-mean-square amplitude of matter density perturbations within a sphere of radius $8\,Mpc h^{-1}$ (being $h$ the dimensionless Hubble parameter) 
is denoted as $\sigma_{8}(z)$ and its evolution is represented by
\begin{equation}\label{Sigma8}
\sigma_{8}(z)=g(z)\sigma_{8}(z=0)\,.
\end{equation}
in where $\sigma_{8}(z=0)$ is the normalization of $\sigma_{8}(z)$ today. Thus, the functions $f$ y $\sigma_{8}$ can be combined to obtain $f\sigma_{8}$ at different 
redshifts. From here, we obtain
\begin{equation}\label{fs8}
f(z)\sigma_{8}(z)=f(z)g(z)\sigma_{8}(z=0).
\end{equation}
The measurements of $f\sigma_{8}$ will be important to constrain different cosmological models.
\section{Current observational data and cosmological constraints.} \label{SectionAllTest}
In this section, we describe how we use the cosmological data currently available to test and constrain the parameter space of our
models proposed.
\subsection{Join Analysis Luminous data set (JLA).} \label{SNIa}
The SNe Ia data sample used in this work is the Join Analysis Luminous data set (JLA) \cite{Betoule2014} composed by $740$ SNe with hight-quality light curves. Here, JLA data
includes several low-redshift samples ($z<0.1$), three samples from the Sloan Digital Sky Survey SDSS-II at $0.05<z<0.4$ and data from Supernova Legacy Survey (SNLS) in
$0.2<z<1.0$.\\
For the JLA data, the observed distance modulus of each SNe is modeled by
\begin{eqnarray}\label{muJLA}
{\mu}^{JLA}_{i}={m}^{*}_{B,i}+\alpha {x_{1,i}}-\beta C_{i}-M_{B}\,,
\end{eqnarray}
in where $1\leq z\leq740$ and the parameters ${m}^{*}_{B}$, $x_{1}$ and $C$ describe the intrinsic variability in the luminosity of the SNe, which are derived from the fitting 
of the light curves. Here, ${m}^{*}_{B}$ is the observed peak magnitude in the rest-frame $B$ band, $x_{1}$ is the stretch measure of the light-curve shape and $C$ is
the color measure for each SNe.\\
On the other hand, the nuisance parameters $\alpha$, $\beta$, $M$ and $dM$ characterize the global properties of the light-curves of the SNe and are estimated 
simultaneously with the cosmological parameters of interest. The parameter $\alpha$ describes the luminosity of the light-curve, $B$ represents the color-luminosity 
relationships, $M$ is the absolute magnitude of the SNe in the rest-frame $B$ band and $dM$ denotes the correction of the absolute magnitude $M$ with host galaxy properties. 
From here, we defined $M_{B}$
\begin{equation}\label{Mb}
M_{B}=\left\{\begin{array}{cl}
M,\, &\mbox{$if$}\,\,\,\rm{M_{stellar}}\,\,\,<10^{10}\rm{M_{\bigodot}}\,,\\
M+dM,\, &\mbox{$if$}\,\,\,\rm{otherwise}\,,           
\end{array}\right.
\end{equation}
where $\rm{M_{stellar}}$ is the host galaxy stellar mass, and $\rm{M_{\bigodot}}$ is the solar mass.\\
The total covariance matrix for this test is denoted as $C_{\bf Betoule}$, and can be written of the following manner
\begin{equation}\label{CovMatirx}
C_{\bf Betoule}=\sigma^{2}_{stat,\,ii}+\sigma^{2}_{stat}+\sigma^{2}_{sys}\,.
\end{equation}
where $\sigma^{2}_{stat,\,ii}$, $\sigma^{2}_{stat}$ and $\sigma^{2}_{sys}$ denote the diagonal part of the statistical covariance matrix, the statistical covariance 
matrix and the systematic covariance matrix, respectively. The details of building of the matrix $C_{\bf Betoule}$ can be found in \cite{Conley2011,Jonsson2010,Betoule2014}.
\begin{widetext}
\begin{eqnarray}\label{CovMatrixDiag}
\sigma^{2}_{stat\,ii}=\sigma^{2}_{{m}^{*}_{B,i}}+\alpha^{2}\sigma^{2}_{x1,i}+\beta^{2}\sigma^{2}_{c,i}+2\alpha Cov({m}_{B,i},{x}_{1,i})
-2\beta Cov({m}_{B,i},{c}_{i})-2\alpha\beta Cov({x}_{1,i},{c}_{i})+\sigma^{2}_{int}+\nonumber\\
+\sigma^{2}_{lensing}+\sigma^{2}_{host\,correction}+\sigma^{2}_{z,i}\biggl[\frac{5(1+z_{i})}{z_{i}(1+0.5z_{i})\ln(10)}\biggr]\,,\qquad\qquad\qquad\qquad\qquad\qquad\qquad\qquad
\end{eqnarray}
\end{widetext}
where the quantities $\sigma^{2}_{{m}^{*}_{B,i}}$, ${\alpha}^{2}{\sigma}^{2}_{x1,i}$ and ${\beta}^{2}{\sigma}^{2}_{c,i}$ are the covariances of ${m}^{*}_{B}$, $x_{1}$ and $C$ for the 
$i$-th SNe, respectively, while $\alpha Cov({m}_{B,i},{x}_{1,i})$, $\beta Cov({m}_{B,i},{c}_{i})$, and $\alpha\beta Cov({x}_{1,i},{c}_{i})$ are the covariances between 
${m}^{*}_{B}$, $x_{1}$ and $C$ for each $i$-th SNe. The terms $\sigma^{2}_{int}$, $\sigma^{2}_{lensing}$ and $\sigma^{2}_{host\,correction}$ account for the uncertainty in 
cosmological redshift due to the following quantities: the peculiar velocities, the variation of magnitudes caused by gravitational lensing and the intrinsic variation in 
SNe magnitude, \cite{Conley2011,Betoule2014}. We follow \cite{Conley2011} in using $c\sigma_{z}=150km{s^{-1}}$ and the prescription suggested by 
J\"onsson for $\sigma_{lensing}=0.055z$ \cite{Jonsson2010}. The values of $\sigma_{host\,correction}$ are compatible with a constant value of $0.106\pm0.006$ 
\cite{Betoule2014}. Furthermore, $\sigma^{2}_{z,i}=0.0005$ denotes the covariance due to a peculiar velocity residual \cite{Conley2011}.\\
On the other hand, the theoretical distance modulus is defined as
\begin{equation}\label{mus}
{\mu}^{\rm{th}}(z,\mathbf{X}) \equiv 5{\log}_{10}\left[\frac{{D_{L}}(z,\mathbf{X})}{\rm{Mpc}}\right]+25\;,
\end{equation}
where the superscript ``$\rm{th}$'' denotes the theoretical prediction for a SNe at a redshift $z$. Likewise, ${D_{L}}(z,\mathbf{X})$ is the luminosity distance, which in a 
FRW cosmology becomes
\begin{equation}\label{luminosity_distance1}
{D}_{L}(z_{hel},z_{CMB},\mathbf{X})=(1+z_{hel})c \int_{0}^{z_{CMB}}\frac{dz'}{H(z',\mathbf{X})}\;,
\end{equation}
where $z_{hel}$ is the heliocentric redshift, $z_{CMB}$ is the CMB rest-frame redshift, ``$c$'' is the speed of the light and $\mathbf{X}$ represents 
the cosmological parameters of the model. Considering that $c=2.9999\times10^{5}km/s$, so we rewrite ${\mu}^{{\rm th}}(z,\mathbf{X})$ as
\begin{eqnarray}\label{mus}
{\mu}^{{\rm th}}(z_{hel},z_{CMB},\mathbf{X})&=&5\log_{10}\biggl[(1+z_{hel}\int_{0}^{z_{CMB}}\frac{dz'}{E(z',\mathbf{X})})\biggr]\nonumber\\
&&+52.385606-5\log_{10}(H_{0})\,.
\end{eqnarray}
Thus, the ${\chi}^{2}$ distribution function for the JLA data is
\begin{equation}\label{X2JLA}
{\chi}_{\bf JLA}^{2}(\mathbf{X})=\left({\Delta{\mu}}_{i}\right)^{t}\left(C^{-1}_{\bf Betoule}\right)_{ij}\left({\Delta{\mu}}_{j}\right)\,,
\end{equation}
where ${\Delta \mu}_{i}={\mu}^{th}_{i}(\mathbf{X})-{\mu}^{JLA}_{i}$ is a column vector of $740$ entries of residuals between the theoretical and distance modulus. 
$C^{-1}_{\bf Betoule}$ is the $740\times740$ covariance matrix for all the observed distance modulus reported in \cite{Betoule2014}, which contains information over both 
systematic and statistical errors.\\
\subsection{RSD data}\label{RSDdata}
RSD data represent a compilation of measurements of the quantity $f(z)\sigma_{8}(z)$ at different redshifts, which were obtained in a model independent way. 
These data are apparent anisotropies (effects) of the galaxy distribution in redshift space due to the differences of the estimates between the redshifts observed distances 
and true distances. They are caused by the component along the line of sight (LOS) of the peculiar velocity of each of the galaxies (recessional speed). Thus, on very small scales 
(a few $Mpc$); especially, in the cores of the clustering of galaxies, the peculiar velocities of galaxies are almost randomly oriented such that the structures of 
the clustering appear elonged along the LOS when they are viewed in the redshift space (the "Finger of God'' effect) \cite{Jackson1972} leading to a damping of the clustering 
of galaxies. By contrast, on large scale (from a few tens of $Mpc$ to $100\,Mpc$) the observations show that the gravitational growth the galaxies tend to fall towards high-density regions and flow 
away from low-density regions such that the galaxy clustering in redshift space is enhanced in the LOS direction in comparison to the transverse direction \cite{Kaiser1987}.\\
The RSD test is an important probe for distinguishing cosmological $DE$ models from standard cosmological models such as $\Lambda$CDM model; namely, different cosmological
models might undergo similar background evolution behavior, but their growth histories of cosmic structures could be distinct in the coupled $DE$ models.\\
In this work, we utilize the most recent growth rate data derived from redshift space distortions on the PSCz, $2$dF, VVDS, $6$dF, $2$MASS, BOSS and WiggleZ galaxy surveys, 
and were collected by Mehrabi et al. (see Table in \cite{Mehrabi2015}). This sample is used to constrain the free parameters of our theoretical models.\\
The standard $\chi^2$ for this data set is defined as \cite{Mehrabi2015}
\begin{equation}\label{X2RSD}
{\chi}^{2}_{RSD}(\mathbf{X}) \equiv \sum_{i=1}^{18}\frac{\left[{f\sigma_{8}}^{th}(\mathbf{X},z_{i})-{f\sigma_{8}}^{obs}(z_{i})\right]^{2}}{{\sigma}^{2}(z_{i})}\;\;,
\end{equation}
where $\sigma(z_{i})$ is the observed $1\sigma$ uncertainty, ${f\sigma_{8}}^{\rm{th}}(\mathbf{X},z_{i})$ and ${f\sigma_{8}}^{obs}(z_{i})$ represent the theoretical and 
observational growth rate, respectively.
\begin{table}
\centering
\begin{tabular}{| c | c | c | c |c | c | c | c |}
 \hline
$z$ & ${f\sigma8}^{obs}$ & $\sigma$ & Refs. &$z$ & ${f\sigma8}^{obs}$ & $\sigma$ & Refs.\\ 
 \hline
$0.020$ & $0.360$& $\pm0.0405$&\cite{Hudson2013} &$0.400$ & $0.419$ & $\pm0.041$ & \cite{Tojeiro2012}  \\
$0.067$ & $0.423$& $\pm0.055$ &\cite{Beutler2012}   &$0.410$ & $0.450$ & $\pm0.040$ & \cite{Blake2011}\\
$0.100$ & $0.370$& $\pm0.130$ &\cite{Feix2015}   &$0.500$ & $0.427$ & $\pm0.043$ & \cite{Tojeiro2012}\\
$0.170$ & $0.510$& $\pm0.060$ &\cite{Percival2004}   &$0.570$ & $0.427$ & $\pm0.066$ & \cite{Reid2012}\\      
$0.220$ & $0.420$& $\pm0.070$ &\cite{Blake2011}   &$0.600$ & $0.430$ & $\pm0.040$ & \cite{Blake2011}\\
$0.250$ & $0.351$& $\pm0.058$ &\cite{Samushia2012}   &$0.600$ & $0.433$ & $\pm0.067$ & \cite{Tojeiro2012}\\ 
$0.300$ & $0.407$& $\pm0.055$ &\cite{Tojeiro2012}   &$0.770$ & $0.490$ & $\pm0.180$ & \cite{Song2009,Guzzo2008}\\
$0.350$ & $0.440$& $\pm0.050$ &\cite{Song2009,Tegmark2006}  &$0.780$ & $0.380$ & $\pm0.040$ & \cite{Blake2011}\\
$0.370$ & $0.460$& $\pm0.038$ &\cite{Samushia2012}   &$0.800$ & $0.470$ & $\pm0.080$ & \cite{delaTorre2013}\\
\hline
\end{tabular}
\caption{Summary of RSD data set \cite{Hudson2013,Beutler2012,Feix2015,Percival2004,Song2009,Tegmark2006,Guzzo2008,
Samushia2012,Blake2011,Tojeiro2012,Reid2012,delaTorre2013}.}
 \label{tableRSD}
\end{table}
\subsection{BAO data sets}\label{BAO}
\subsubsection{$\bf{\rm{BAO}}$\,\,$\bf{\rm{I}}$\, data}\label{BAOI}
In this work, we make use of six different BAO galaxies clustering observations from six-degree-field galaxy survey ($6$dFGRS) \cite{Hinshaw2013,Beutler2011}, the Sloan 
Digital Sky Survey (SDSS) Data Releases (DR) (such as SDSS-DR$7$ \cite{Blake2011,Ross2015,Percival2010,Kazin2010,Padmanabhan2012,Chuang2013a}, SDSS-DR$9$ 
\cite{Chuang2013b,Kazin2014} and SDSS-DR$11$ \cite{Anderson2014a}, respectively.), the Wiggle $Z$ dark energy survey \cite{Blake2011} and the $Ly\alpha$ forest
measurements from Baryon Oscillation Spectroscopic Data Release $11$ (BOSS $11$) \cite{Anderson2014a,Debulac2015,FontRibera2014}.
Eisenstein et al. \cite{Eisenstein1998} and Percival et al. \cite{Percival2010} constructed an effective distance ratio $D_{v}(z)$, which encodes the visual 
distortion of a spherical object due to the non-Euclidianity of a FRW spacetime. It is defined as
\begin{eqnarray}\label{Dv}
D_{v}(z,\mathbf{X})&\equiv&\frac{1}{H_{0}}\left[(1+z)^{2}{D_{A}}^{2}(z)\frac{cz}{E(z)}\right]^{1/3},\nonumber\\
&=&\frac{c}{H_{0}}\left[\left(\int_{0}^{z}\frac{dz'}{E(z',\mathbf{X})}\right)^{2}\frac{z}{E(z,\mathbf{X})}\right]^{1/3}
\end{eqnarray}
where $D_{A}(z)$ is the proper (not comoving) angular diameter distance, which has the following definition
\begin{eqnarray} \label{DA}
D_{A}(z,\mathbf{X}) &\equiv& \frac{c}{(1+z)}{\int}^{z}_{0}\frac{dz'}{H(z',\mathbf{X})}\;.
\end{eqnarray}
The comoving sound horizon size is defined by
\begin{equation} \label{hsd} 
r_{s}(a)\equiv c\int^{a}_{0}\frac{c_{s}(a')da'}{{a'}^{2}H(a')}\;\;,
\end{equation}
being $c_{s}(a)$ the sound speed of the photon-baryon fluid
\begin{equation} \label{vsd}
c_{s}^{2}(a) \equiv \frac{\delta P}{\delta \rho}=\frac{1}{3}\left[\frac{1}{1+(3\Omega_{b}/4\Omega_{r})a}\right]\;\;.
\end{equation}
Considering Eqs. (\ref{hsd}) and (\ref{vsd}) for a $z$, we have   
\begin{equation}\label{rs}
r_{s}(z)=\frac{c}{\sqrt 3}{\int}^{1/(1+z)}_{0}\frac{da}{a^{2}H(a)\sqrt{1+(3\Omega_{b,0}/4\Omega_{\gamma,0})a}}\;\;\;,
\end{equation}
where $\Omega_{b,0}$ and $\Omega_{\gamma,0}$ are the present-day baryon and photon density parameters, respectively. In this paper, we have fixed 
$\Omega_{\gamma,0}=2.469\times10^{-5}h^{-2}$, $\Omega_{b,0}=0.02230h^{-2}$, and $\Omega_{r,0}=\Omega_{\gamma,0}(1+0.2271N_{eff})$, where $N_{eff}$ represents the effective 
number of neutrino species (here, $\Omega_{b,0}$, $\Omega_{r,0}$ and the standard value, $N_{eff}=3.04\pm0.18$ were chosen from Table $4$ in \cite{Planck2015}).\\
The epoch in which the baryons were released from photons is denoted as, $z_{d}$, and can be determined by using the following fitting formula \cite{Eisenstein1998}:
\begin{equation}\label{zd}
z_{d}=\frac{1291(\Omega_{M,0}h^{2})^{0.251}}{1+0.659(\Omega_{M,0}h^{2})^{0.828}}\left(1+b_{1}(\Omega_{b,0}h^{2})^{b_{2}}\right)\;,
\end{equation}
where $\Omega_{M,0}=\Omega_{DM,0}+\Omega_{b,0}$, and
\begin{eqnarray*} 
b_{1}&=&0.313(\Omega_{M,0}h^{2})^{-0.419}\left[1+ 0.607(\Omega_{M,0}h^{2})^{0.674}\right]\;,\\
b_{2}&=&0.238(\Omega_{M,0}h^{2})^{0.223}\;.
\end{eqnarray*}
The peak position of the BAO depends of the distance radios $d_{z}$ at different redshifts, which were obtained from the surveys already listed in Table \ref{tableBAOI}.\\
\begin{equation} \label{dvalues}
d_{z}(\mathbf{X})=\frac{r_{s}(z_{d})}{D_{V}(z,\mathbf{X})}\;,
\end{equation}
where $r_{s}(z_{d}, \mathbf{X})$ is the comoving sound horizon size at the baryon drag epoch. From the data showed in Table \ref{tableBAOI}, we can build the $\chi^{2}$ for 
the BAO\,$\rm{I}$ data
\begin{table}
\centering
\begin{tabular}{| c | c | c | c |c | c |c | c |}
 \hline
 $z$ & $d_{z}^{obs}$ & $\sigma_{z}$& Refs. & $z$ & $d_{z}^{obs}$ & $\sigma$& Refs.\\
 \hline
$0.106$ & $0.3360$ &$\pm0.0150$ &\cite{Hinshaw2013,Beutler2011} &$0.350$ & $0.1161$ &$\pm0.0146$&\cite{Chuang2013a}\\
$0.150$ & $0.2232$ &$\pm0.0084$ &\cite{Ross2015} &$0.440$ & $0.0916$ &$\pm0.0071$&\cite{Blake2011}\\
$0.200$ & $0.1905$ &$\pm0.0061$ &\cite{Percival2010,Blake2011} &$0.570$ & $0.0739$ &$\pm0.0043$&\cite{Chuang2013b}\\
$0.275$ & $0.1390$ &$\pm0.0037$ &\cite{Percival2010} &$0.570$ & $0.0726$ &$\pm0.0014$&\cite{Anderson2014a}\\
$0.278$ & $0.1394$ &$\pm0.0049$ &\cite{Kazin2010} &$0.600$ & $0.0726$ &$\pm0.0034$&\cite{Blake2011}\\
$0.314$ & $0.1239$ &$\pm0.0033$ &\cite{Blake2011} &$0.730$ & $0.0592$ &$\pm0.0032$&\cite{Blake2011}\\
$0.320$ & $0.1181$ &$\pm0.0026$ &\cite{Anderson2014a}&$2.340$ & $0.0320$ &$\pm0.0021$&\cite{Debulac2015}\\
$0.350$ & $0.1097$ &$\pm0.0036$ &\cite{Percival2010,Blake2011} &$2.360$ & $0.0329$ &$\pm0.0017$&\cite{FontRibera2014}\\
$0.350$ & $0.1126$ &$\pm0.0022$ &\cite{Padmanabhan2012}         &         &       \\ 
 \hline
\end{tabular}
\caption{Summary of BAO data set \cite{Percival2010,Hinshaw2013,Beutler2011,Ross2015,Blake2011,Kazin2010,Padmanabhan2012,Chuang2013a,Chuang2013b,Anderson2014a,Debulac2015,FontRibera2014}.}
 \label{tableBAOI}
\end{table}
\begin{equation}\label{X2BAOI}
\chi_{\bf{\bm{BAO}}\,\rm{I}}^{2}(\mathbf{X})=\sum_{i=1}^{17}\left(\frac{d_{z}^{th}(\mathbf{X},z_{i})-d_{z}^{obs}(\mathbf{X},z_{i})}{\sigma(\mathbf{X},z_{i})}\right)^{2}\;\;.
\end{equation}
\subsubsection{$\bf{\rm{BAO}}$\,\,$\bf{\rm{II}}$\,data}\label{BAOII}
From BOSS DR $9$ CMASS sample, Chuang et al. in \cite{Chuang2013b} analyzed the shape of the monopole and quadrupole from the two-dimensional two-points 
correlation function $2$d$2$pCF of galaxies and measured simultaneously $H(z)$, $D_{A}(z)$, $\Omega_{m}h^{2}$ and $f(z)\sigma_{8}(z)$ at the effective redshift $z=0.57$. These results 
were $H(0.57)=87.6\begin{array}{cl}+6.7\\-6.8\end{array}$, $D_{A}(0.57)=1396\pm 73$, $\Omega_{m}h^{2}(0.57)=0.126\begin{array}{cl}+0.008\\-0.010\end{array}$ and 
$f(0.57)\sigma_{8}(0.57)=0.428\pm 0.066$. The units for $H$ and $D_{A}$ are $Kms^{-1}Mpc^{-1}$ and $Mpc$, respectively. Here, ${\Delta A}_{i}=A^{th}_{i}(\mathbf{X})-A^{obs}_{i}$ 
is a column vector defined as
\begin{equation}\label{bossdr9}
{\Delta A}_{i}=\left(\begin{array}{rl}
H(0.57)-87.6\\
D_{A}(0.57)-1396\\
\Omega_{m}h^{2}(0.57)-0.126\\
f(0.57)\sigma_{8}(0.57)-0.428\\
\end{array}\right)\;,
\end{equation}
Then, the $\chi^{2}$ function for the BAO $\rm{II}$ data is given by 
\begin{equation}\label{X2BAOII}
\chi_{\bf BAO\,\rm{II}}^{2}(\mathbf{X})=\left({\Delta A}_{i}\right)^{t}\left(C^{-1}_{\bf BAO\,\rm{IV}}\right)_{ij}\left({\Delta A}_{j}\right),
\end{equation}
where the covariance matrix of measurements is listed in Eq. ($26$) of \cite{Chuang2013b}
\begin{equation}\label{BAOII}
C^{-1}_{\bf{\rm{BAO}}\,\bf{\rm{II}}}=\left(\begin{array}{lccr}
 +0.03850 \,\,-0.0011410 \,-13.53 \,-1.2710\\
 -0.001141 +0.0008662   \,+3.354 \,-0.3059\\
 -13.530  \,\,\,\,\,+3.3540\quad\,\,\,\,\,+19370\,-770.0\\
 -1.2710  \,\,\,\,\,-0.30590 \quad\,-770.0\,\,+411.3
\end{array} \right)\;.
\end{equation}
where ``t'' denotes its transpose.\\
\subsubsection{$\bf{\rm{BAO}}$\,\,$\bf{\rm{III}}$\, data}\label{BAOIII}
Using SDSS DR $7$ sample Hemantha et al \cite{Hemantha2014}, proposed a new method to constrain $H(z)$ and $D_{A}(z)$ simultaneously from the two-dimensional matter power 
spectrum $2$dMPS without assuming a dark energy model or a flat universe. The values obtained at the effective redshift $z=0.35$ were $H(0.35)=81.3\pm3.8 Kms^{-1}Mpc^{-1}$, 
$D_{A}(0.35)=1037\pm44 Mpc$ and $\Omega_{m}h^{2}(0.35)=0.1268\pm0.0085$. They defined a column vector ${\Delta B}_{i}=B^{th}_{i}(\mathbf{X})-B^{obs}_{i}$ as
\begin{equation}
B^{th}_{i}(\mathbf{X})-B^{obs}_{i}=\left(\begin{array}{rl}                               
 H(0.35,\mathbf{X})-81.3\\
 D_{A}(0.35,\mathbf{X})-1037.0\\
 \Omega_{M}h^{2}(0.35,\mathbf{X})-0.1268\\
\end{array}\right)\;.
\end{equation}
The covariance matrix for the set of cosmological parameters under consideration was
\begin{equation}\label{BAOIII}
C^{-1}_{\bf BAO\,\rm{III}}=\left(\begin{array}{lcr}           
+0.00007225 \, -0.169606 \,+0.01594328\\
-0.1696090  \,\,\,+1936.0\,\,\quad+67.030480 \\
+0.01594328 \, +67.03048\,+14.440\\
\end{array} \right)\;.
\end{equation}
The $\chi^{2}$ function for the BAO $\rm{III}$ data set is written as  
\begin{equation}\label{X2BAOIII}
\chi_{\bf{\rm{BAO}}\,\bf{\rm{III}}}^{2}(\mathbf{X})=\left({\Delta B}_{i}\right)^{t}\left(C^{-1}_{\bf BAO \rm{III}}\right)_{ij}\left({\Delta B}_{j}\right),
\end{equation}
where ``t'' denotes its transpose.\\
\subsubsection{$\bf{\rm{BAO}}$\,\,$\bf{\rm{IV}}$\, data}\label{BAOIV}
In all the catalogs of galaxies, the positions of them are given in terms of angular positions and redshifts. In order to measure clustering of galaxies, we need 
to convert angular positions and redshifts of galaxies into physical positions, just for that we must use a fiducial cosmological model. These physical distances will depend
on the chosen fiducial model. If the fiducial cosmology is significantly different from the real (true) cosmology, then this difference will induce any measured anisotropy, 
and should be used to constrain the true cosmology of the universe. This is known as the AP test. This signal affirms that if an astrophysical structure is spherically symmetric
or isotropic, then it should possess equal comoving sizes, $r_{s}$, in parallel and transverse dimensions to the LOS \cite{Alcock1979}. Thus, the comoving diameter 
of a spherical object $r_{s}$ at redshift $z$ is related to its angular size ($\Delta \theta$) on the sky by $\Delta \theta=r_{s}/[(1+z)D_{A}]$, which is known as observed 
transverse dimension, whilst the parallel dimension, $r_{s}$ can also be related to the redshift difference by $\Delta z=r_{s}H(z)/c$. Furthermore, any difference between 
the relative values of $z$, $H(z)$ and $D_{A}$ of an astrophysical structure in the fiducial cosmology and in the true cosmology, will manifest as 
anisotropies along the LOS. The parallel and transverse dimensions can be conveniently combined in a single parameter $F_{AP}(z)$, defined as
\begin{equation}\label{Fap}
F_{AP}(z)=\frac{\Delta z}{\Delta \theta}=(1+z)D_{A}(z)\left(H(z)/c\right)\,, 
\end{equation}
where $F_{AP}(z)$ is known as the AP distortion parameter.\\
Measuring this parameter we can obtain accurate estimates of the angular distance $D_{A}(z)$ and Hubble parameter $H(z)$; likewise, we could break the degeneracy between them. 
For this reason, $F_{AP}$ can also be used to constrain the properties of the $DE$ \cite{Seo2008}.\\
It is convenient to report the results of the $BAO$ peak, the AP test and the RSD effect, as joint measurements of $d_{z}(z_{eff})$, $F_{AP}(z_{eff})$ and
$f(z_{eff})\sigma_{8}(z_{eff})$, where $z_{eff}$ is an effective redshift. This joint measurements can be used to constrain cosmological parameters, and also, to distinguish 
different $DE$ models. Then, we define a vector $V$ with all these measurements at $z_{eff}=0.57$, which can be built as \cite{Anderson2014a, Battye2015, Samushia2014} \\
\begin{equation}\label{bossdr11}
{\Delta V}_{i}=V^{th}_{i}(\mathbf{X})-V^{obs}_{i}=\left(\begin{array}{rl}
d_{z}(z_{eff})-13.880\\
F_{AP}(z_{eff})-0.683\\
f(z_{eff})\sigma_{8}(z_{eff})-0.422\\
\end{array}\right)\;,
\end{equation}
The $\chi^{2}$ function for this data set is fixed as
\begin{equation}\label{X2BAOIV}
\chi_{\bf BAO\,\rm{IV}}^{2}(\mathbf{X})=\left({\Delta V}_{i}\right)^{t}\left(C^{-1}_{\bf BAO\,\rm{IV}}\right)_{ij}\left({\Delta V}_{j}\right)\,,
\end{equation}
where the covariance matrix of measurements is listed in Eq. ($1.3$) of \cite{Battye2015}
\begin{equation}\label{BAOIV}
C^{-1}_{\bf{\rm{BAO}}\,\bf{\rm{IV}}}=\left(\begin{array}{lcr}           
+31.032  \,+77.773 \,-16.796\\
+77.773 \, +2687.7 \,-1475.9\\
-16.796 \, -1475.9 \,+1323.0\\
\end{array} \right)\;.
\end{equation}
Considering the Eqs. (\ref{X2BAOI}), (\ref{X2BAOII}), (\ref{X2BAOIII}) and (\ref{X2BAOIV}), we construct the total $\chi_{\bf BAO}^{2}$ for all the BAO data sets
\begin{equation}\label{X2Total}
 {{{\rm{\bf{\chi}}^{2}}}}_{\bf BAO}=\chi_{\bf{BAO\,\rm{I}}}^{2}+\chi_{\bf{BAO\,\rm{II}}}^{2}+\chi_{\bf{BAO\,\rm{III}}}^{2}+\chi_{\bf{BAO\,\rm{IV}}}^{2}\,,
\end{equation}
\subsection{CMB data set} \label{CMB}
The JLA (SNe Ia) and BAO data sets contain information about the universe at low redshifts, we now include Planck $2015$ data \cite{Planck2015} to probe the
entire expansion history up to the last scattering surface. The shift parameter ${\rm {\bf R}}$ is provided by \cite{Bond-Tegmark1997}
\begin{eqnarray}\label{Shiftparameter}
{\rm {\bf R}}(z_{*},\mathbf{X})\equiv \frac{H_{0}}{c}\sqrt{\Omega_{M,0}}(1+z_{*})D_{A}(z_{*},\mathbf{X}),\nonumber\\
=\sqrt{\Omega_{M,0}}{\int}^{z_{*}}_{0}\frac{d\tilde{y}}{E(\tilde{y})}\,,\hspace{2cm}
\end{eqnarray}
where the distance $D_{A}$ and $E(\tilde{y})$ are given by Eqs. (\ref{DA}) and (\ref{hubble}), respectively. 
Moreover, the redshift $z_{*}$ (the decoupling epoch of photons) is obtained using the following fitting function \cite{Hu-Sugiyama1996}
\begin{equation}\label{Redshift_decoupling}
{z}_{*}=1048\biggl[1+0.00124({\Omega}_{b,0}h^{2})^{-0.738}\biggr]\biggl[1+{g}_{1}({\Omega}_{M,0}h^{2})^{{g}_{2}}\biggr]\;,\;\;
\end{equation}
where $\Omega_{M,0}=\Omega_{DM,0}+\Omega_{b,0}$, and 
\begin{equation}\label{g1g2}
g_{1}=\frac{0.0783(\Omega_{b,0}h^{2})^{-0.238}}{1+39.5(\Omega_{b,0}h^{2})^{0.763}}\;,\hspace{0.3cm}g_{2}=\frac{0.560}{1+21.1(\Omega_{b,0}h^{2})^{1.81}}.
\end{equation}
An angular scale $l_{A}$ for the sound horizon at decoupling epoch is defined as
\begin{equation}\label{Acoustic_scale}
l_{A}(\mathbf{X})\equiv(1+z_{*})\frac{\pi D_{A}(z_{*},\mathbf{X})}{r_{s}(z_{*},\mathbf{X})}\,,\hspace{1cm}
\end{equation}
where $r_{s}(z_{*},\mathbf{X})$ is the comoving sound horizon at $z_{*}$, and is given by Eq. (\ref{rs}).
Then, following \cite{Planck2015,Neveu2016}, the $\chi^{2}$ for the CMB data is
\begin{equation}\label{X2CMB}
\chi_{\bf CMB}^{2}(\mathbf{X})=\left({\Delta x}_{i}\right)^{t}\left(C^{-1}_{\bf CMB}\right)_{ij}\left({\Delta x}_{j}\right)\,,
\end{equation}
where ${\Delta x}_{i}=x^{th}_{i}(\mathbf{X})-x^{obs}_{i}$ is a column vector 
\begin{equation}
x^{th}_{i}(\mathbf{X})-x^{obs}_{i}=\left(\begin{array}{cc}
 l_{A}(z_{*})-301.7870\\
 R(z_{*})-1.7492\\
 \;\;z_{*}-1089.990\\
\end{array}\right)\;,
\end{equation}
``t'' denotes its transpose and $(C^{-1}_{\bf CMB})_{ij}$ is the inverse covariance matrix \cite{Neveu2016}
given by
\begin{equation}\label{MatrixCMB}
C^{-1}_{\bf CMB}\equiv\left(
\begin{array}{ccc}
+162.48&-1529.4&+2.0688\\
-1529.4&+207232&-2866.8\\
+2.0688&-2866.8&+53.572\\
\end{array}\right)\;.
\end{equation}
The errors for the CMB data are contained in $C^{-1}_{\bf CMB}$.
\begin{table*}[!htb]
\begin{tabular}{>{\centering\arraybackslash}m{7.8cm} >{\arraybackslash}m{10cm}}
{\begin{tabular}{| c  | c  | c  | c  | c  | c  | c  | c | }
\hline
 $z$   & $H(z)$ &  $1\sigma$& Refs. & $z$   & $H(z)$ &  $1\sigma$& Refs. \\
\hline
$0.070$&  $69.0$&  $\pm19.6$&\cite{Zhang2014}    & $0.570$&  $96.8$&  $\pm3.40$&\cite{Anderson2014a}\\
$0.090$&  $69.0$&  $\pm12.0$&\cite{Simon2005}    & $0.593$& $104.0$&  $\pm13.0$&\cite{Moresco2012}\\
$0.120$&  $68.6$&  $\pm26.2$&\cite{Zhang2014}    & $0.600$&  $87.9$&  $\pm6.1$ &\cite{Blake2012}\\
$0.170$&  $83.0$&  $\pm8.0$&\cite{Simon2005}     & $0.680$&  $92.0$&  $\pm8.0$ &\cite{Moresco2012}\\
$0.179$&  $75.0$&  $\pm4.0$&\cite{Moresco2012}   & $0.730$&  $97.3$&  $\pm7.0$ &\cite{Blake2012}\\
$0.199$&  $75.0$&  $\pm5.0$&\cite{Moresco2012}   & $0.781$& $105.0$&  $\pm12.0$&\cite{Moresco2012}\\
$0.200$&  $72.9$&  $\pm29.6$&\cite{Zhang2014}    & $0.875$& $125.0$&  $\pm17.0$&\cite{Moresco2012}\\
$0.240$&  $79.69$& $\pm2.99$&\cite{Gastanaga2009}& $0.880$&  $90.0$&  $\pm40.0$&\cite{Stern2010}\\
$0.270$&  $77.0$&  $\pm14.0$&\cite{Simon2005}    & $0.900$& $117.0$&  $\pm23.0$&\cite{Simon2005}\\
$0.280$&  $88.8$&  $\pm36.6$&\cite{Zhang2014}    & $1.037$& $154.0$&  $\pm20.0$&\cite{Gastanaga2009}\\
$0.300$&  $81.7$&  $\pm6.22$&\cite{Oka2014}      & $1.300$& $168.0$&  $\pm17.0$&\cite{Simon2005}\\
$0.340$&  $83.8$&  $\pm3.66$&\cite{Gastanaga2009}& $1.363$& $160.0$&  $\pm33.6$&\cite{Moresco2015}\\
$0.350$&  $82.7$&  $\pm9.1$& \cite{Chuang2013a}  & $1.430$& $177.0$&  $\pm18.0$&\cite{Simon2005}\\
$0.352$&  $83.0$&  $\pm14.0$&\cite{Moresco2012}  & $1.530$& $140.0$&  $\pm14.0$&\cite{Simon2005}\\
$0.400$&  $95.0$&  $\pm17.0$&\cite{Simon2005}    & $1.750$& $202.0$&  $\pm40.0$&\cite{Simon2005}\\
$0.430$&  $86.45$& $\pm3.97$&\cite{Gastanaga2009}& $1.965$& $186.5$&  $\pm50.4$&\cite{Moresco2015}\\
$0.440$&  $82.6$&  $\pm7.8$&\cite{Blake2012}     & $2.300$& $224.0$&  $\pm8.6$ &\cite{Busca2013}\\
$0.480$&  $97.0$&  $\pm62.0$&\cite{Stern2010}    & $2.340$& $222.0$&  $\pm8.5$ &\cite{Debulac2015}\\
$0.570$&  $87.6$&  $\pm7.80$&\cite{Chuang2013b}  & $2.360$& $226.0$&  $\pm9.3$ &\cite{FontRibera2014}\\

\hline\end{tabular}
\caption{Shows the observational $H(z)$ data \cite{Chuang2013a,Chuang2013b,Anderson2014a,Debulac2015,FontRibera2014,Zhang2014,Simon2005,Moresco2012,Gastanaga2009,
Oka2014,Blake2012,Stern2010,Moresco2015,Busca2013}}
\label{tableOHD}} & 
{\begin{tabular}{|c|@{\extracolsep{0mm}\ }c@{ }|}
\hline
Parameters&Constant Priors\\[0.2mm]
\hline
${\lambda}_{0}$&$[-1.5\times10^{+2},+1.5\times10^{+2}]$\\[0.2mm]
${\lambda}_{1}$&$[-1.5\times10^{+2},+1.5\times10^{+2}]$\\[0.2mm]
${\lambda}_{2}$&$[-1.5\times10^{+1},+1.5\times10^{+1}]$\\[0.2mm]
$\omega_{0}$&$[-2.0,-0.3]$\\[0.2mm]
$\omega_{1}$&$[-1.0,+1.0]$\\[0.2mm]
$\omega_{2}$&$[-2.0,+0.1]$\\[0.2mm]
$\Omega_{DM,0}$&$[0,0.7]$\\[0.2mm]
$H_{0}(kms^{-1}{Mpc}^{-1})$&$[20,120]$\\[0.2mm]
$\alpha$&$[-0.2,+0.5]$\\[0.2mm]
$\beta$&$[+2.1,+3.8]$\\[0.2mm]
$M$&$[-20,-17]$\\[0.2mm]
$dM$&$[-1.0,+1.0]$\\[0.2mm]
$\gamma_{0}$&$[+0.2,+1.2]$\\[0.2mm]
$\sigma_{80}$&$[0,+1.65]$\\[0.2mm]
\hline
\end{tabular} 
\caption{Shows the priors on the parameter space.\hspace{2.7cm}}\label{Priors}}
\end{tabular}
\end{table*}  
\subsection{Observational Hubble data (H)}\label{OHD}
Recently G. S. Sharov \cite{Sharov2015} compiled a list of $38$ independent measurements of the Hubble parameter at different redshitfs, and used these measurements to 
constrain different cosmological models (see Table $\rm{III}$ in \cite{Sharov2015}). These data points were derived from two different methods: The first one includes 
twenty-five points, which were obtained from differential age $dt$ for passively evolving galaxies with redshifts $dz$, 
(see \cite{Zhang2014,Simon2005,Moresco2012,Stern2010,Moresco2015})
\begin{equation}\label{Hubble}
{\bf H}(z)=-\frac{1}{1+z}\frac{dz}{dt}\;. 
\end{equation}
The second one contains $13$ data points \cite{Chuang2013a,Chuang2013b,Anderson2014a,Debulac2015,FontRibera2014,Gastanaga2009,Oka2014,Blake2012,Busca2013}, and were 
determined by using the two-point correlation of Sloan Digital Sky Survey. Here, the BAO peak position was considered as a standard ruler in the radial direction.\\
The $\chi^2_{H}$ function for this data set is
\begin{equation}\label{X2OHD}
\chi^2_{\,\bf{\bm{H}}}(\mathbf{X})\equiv\sum_{i=1}^{38}\frac{\left[H^{\rm {th}}(\mathbf{X},z_{i},)-H^{obs}(z_{i})\right]^2}{\sigma^2(z_{i})}\;\;,
\end{equation}
where $\mathbf{X}$ represents the parameters of the model, $H^{{\rm th}}$ is the theoretical value for the Hubble parameter, $H^{obs}$ is the observed value, 
$\sigma(z_{i})$ is the standard deviation measurement uncertainty, and the summation is over the $38$ observational Hubble data at $z_{i}$. This test has been 
already used to constrain some models in \cite{Sharov2015}.\\
Therefore, the best fitted parameters are obtained by minimizing the following total function $\chi^2$, 
\begin{equation}\label{TotalChi}
{{{\rm {\bf{\chi}}^{2}}}}={{{\rm{\bf{\chi}}^{2}}}}_{\bf JLA}+{{{\rm {\bf {\chi}}^{2}}}}_{\bf{RSD}}+{{{\rm {\bf {\chi}}^{2}}}}_{\bf BAO}+{{{\rm {\bf {\chi}}^{2}}}}_{\bf CMB}+
{{{\rm {\bf {\chi}}^{2}}}}_{\bf H}\;\;.
\end{equation}
By means of this relation, we can construct the total probability density function, ${{\rm {\bf pdf}}}$ as 
\begin{equation}\label{TotalexpChi}
{{\rm {\bf pdf}}}(\mathbf{X})=\rm{A}{{\rm e}}^{-{{\chi}}^{2}/2}\;\;.
\end{equation}
where $\rm A$ is a integration constant.
\subsection{Constant Priors}\label{Values}
In this work, we have assumed that baryonic matter ($b$) and radiation ($r$) are not coupled to $DE$ or $DM$, which are separately conserved \cite{Koyama2009-Brax2010}. 
In this regard, we believe that the intensity of the interaction, ${\rm I}_{\rm Q}$, is not affected by the values of $\Omega_{b,0}$ and $\Omega_{r,0}$, respectively. 
Thus, in this paper, we fixed: $\Omega_{\gamma,0}=2.469\times10^{-5}h^{-2}$ and $\Omega_{b,0}=0.02230h^{-2}$, given by Planck 2015 data \cite{Planck2015}. Using 
these assumptions, in each of our models, we will construct a ${{\rm {\bf pdf}}}$ function for them. The priors on the parameters space are given in Table \ref{Priors}, 
and were used in all our observational tests. From they we will compute the best fitting parameters.
\begin{figure*}[!htb]
 \includegraphics[width=8.9cm,height=8.6cm]{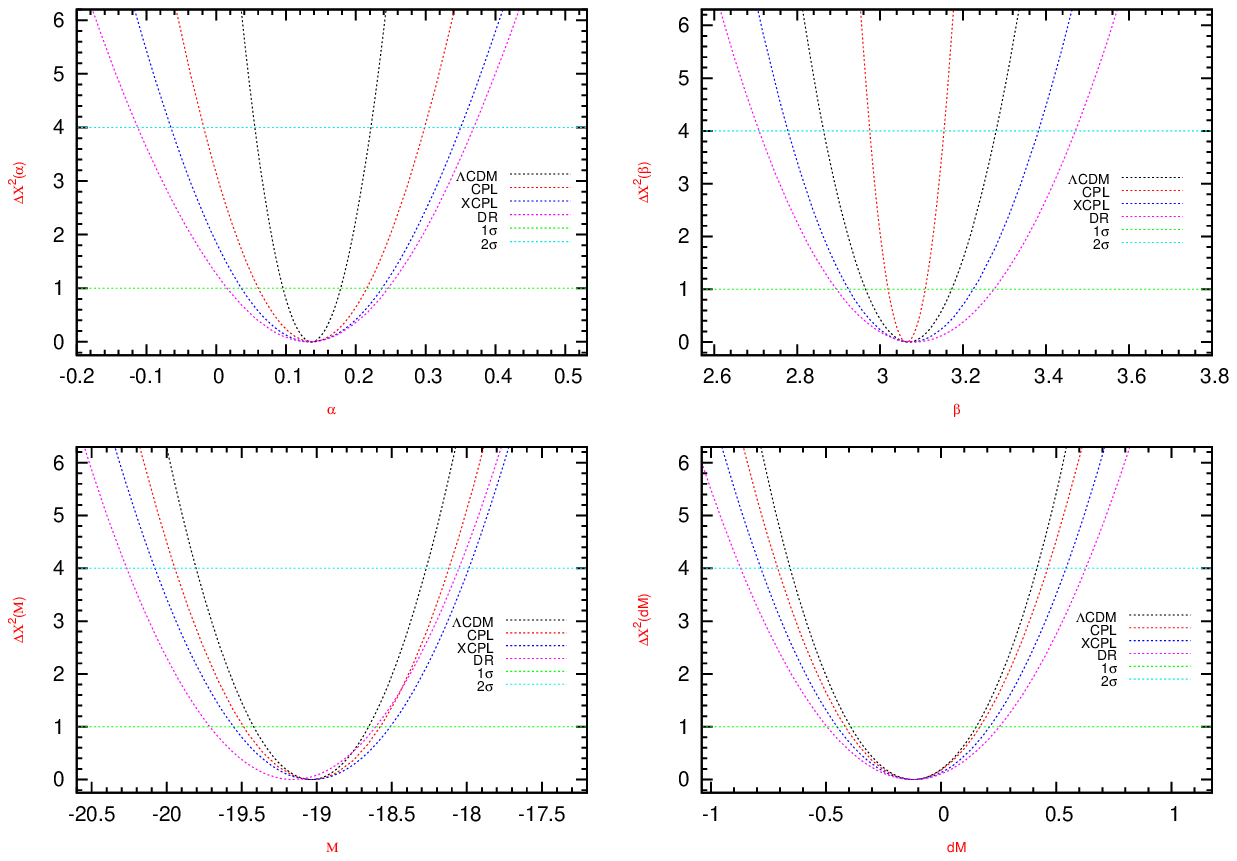}\includegraphics[width=8.9cm,height=8.6cm]{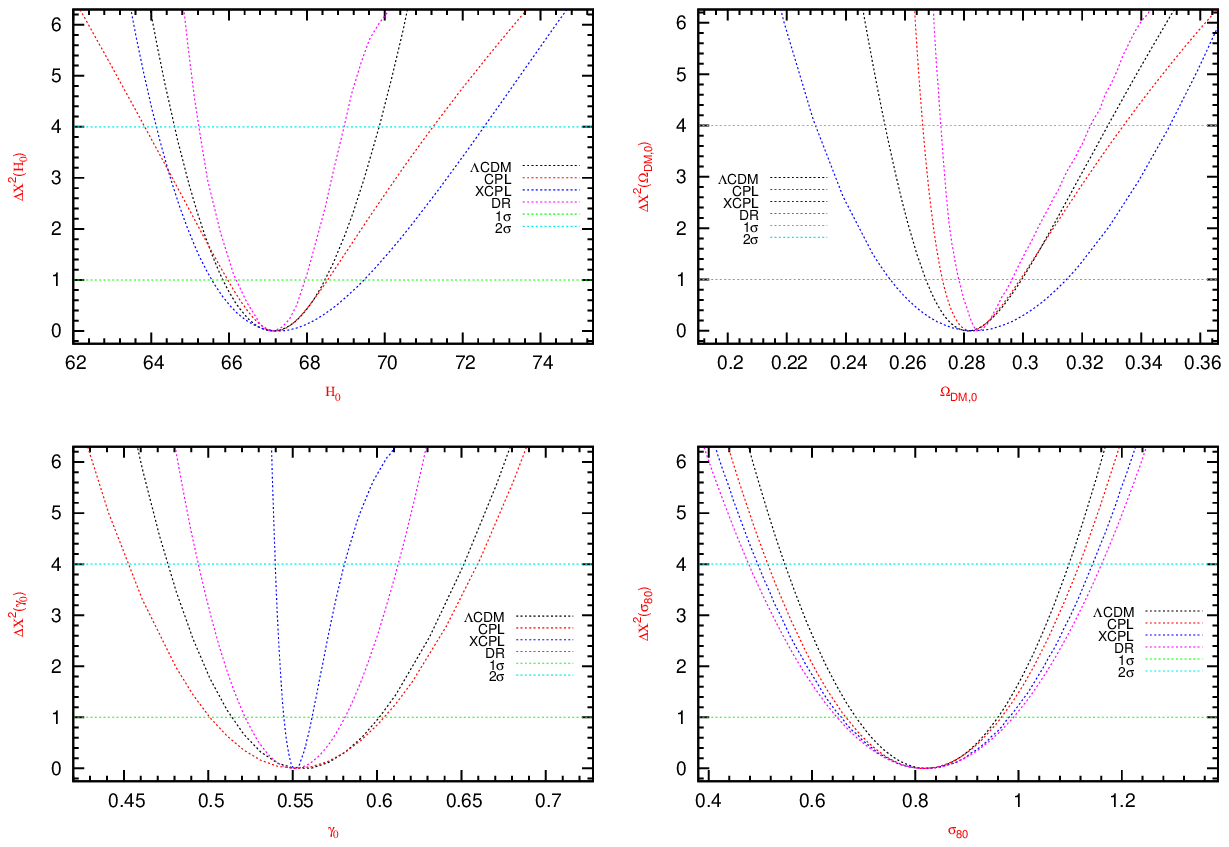}\\
 \includegraphics[width=8.9cm,height=8.6cm]{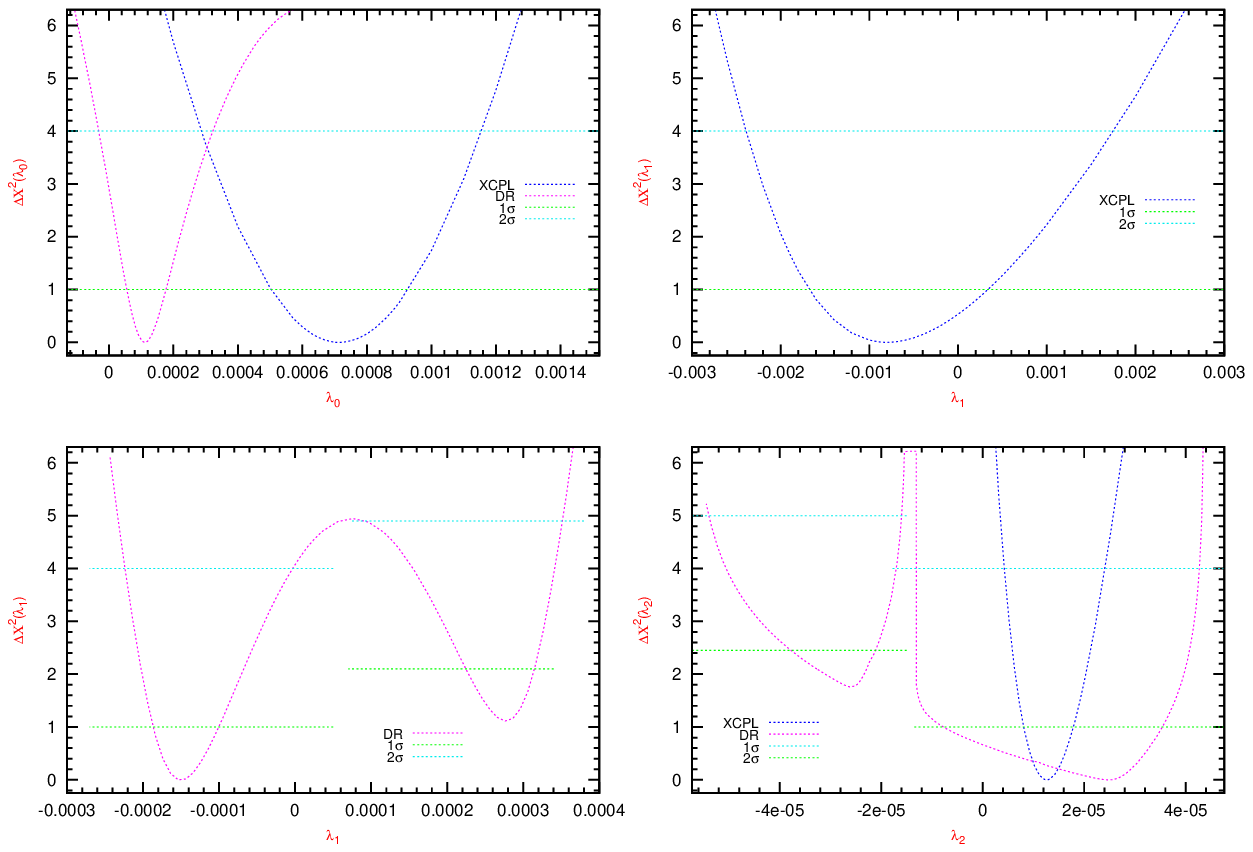}\includegraphics[width=8.9cm,height=8.6cm]{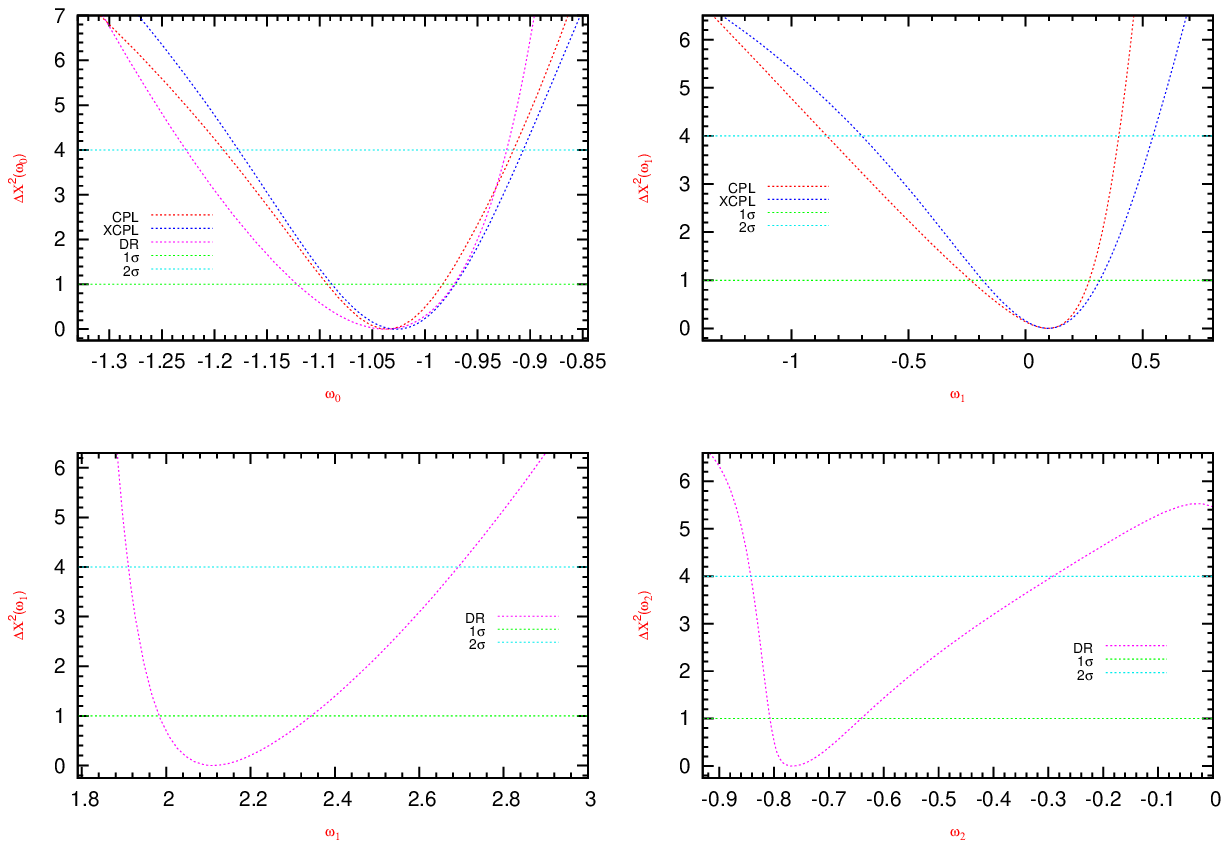}\\
 \caption{(color online) Displays the one-dimension probability contours for all the parameters worked and their constraints at $1\sigma$ and $2\sigma$, respectively. 
 Moreover, we consider that $\Delta{\chi^{2}}=\chi^{2}-{\chi^{2}_{min}}$\,.}\label{Contours}
\end{figure*} 
\begingroup
\squeezetable
\begin{table*}[!hbtp]
\centering
\caption{Shows the best fitting cosmological parameters for each model and their constraints at $1\sigma$ and $2\sigma$ obtained from an analysis of 
Union JLA+RSD+BAO+CMB+H data sets.}\label{Bestfits}
\begin{tabular}{| c | c | c | c |}
\hline
 Parameters&$\Lambda$CDM&CPL&XCPL\\
 \hline\hline 
${\lambda}_{0}$&$N/A$&$N/A$&${+7.0\times10^{-4}}^{+2.2609\times10^{-4}+4.5085\times10^{-4}}_{-1.9541\times10^{-4}-4.1348\times10^{-4}}$\\[0.4mm]
${\lambda}_{1}$&$N/A$&$N/A$&${-8.0\times10^{-4}}^{+11.4167\times10^{-4}+25.3575\times10^{-4}}_{-8.7161\times10^{-4}-15.9322\times10^{-4}}$\\[0.4mm]
${\lambda}_{2}$&$N/A$&$N/A$&${+1.27\times10^{-5}}^{+0.5128\times10^{-5}+1.1332\times10^{-5}}_{-0.4640\times10^{-5}-0.8529\times10^{-5}}$\\[0.6mm]
$\omega_{0}$&$-1.0$&${-1.0323}^{+0.0489+0.1165}_{-0.0605-0.1586}$&${-1.0271}^{+0.0563+0.1208}_{-0.0610-0.1497}$\\[0.4mm]
$\omega_{1}$&$N/A$&${+0.0952}^{+0.1757+0.3024}_{-0.3267-0.9446}$&${+0.0950}^{+0.2218+0.4488}_{-0.2827-0.7960}$\\[0.4mm]
$\omega_{2}$&$N/A$&$N/A$&$N/A$\\[0.4mm]
$\Omega_{DM,0}$&${+0.2810}^{+0.0185+0.0476}_{-0.0138-0.0279}$&${+0.2814}^{+0.0176+0.0528}_{-0.0089-0.0154}$&${+0.2840}^{+0.0308+0.0659}_{-0.0290-0.0542}$\\[0.4mm]
$H_{0}(kms^{-1}{Mpc}^{-1})$&${+67.170}^{+1.274+2.694}_{-1.3079-2.5666}$&${+67.19}^{+1.3508+4.0403}_{-1.2203-3.3504}$&${+67.20}^{+2.2767+5.3535}_{-1.6607-3.0699}$\\[0.4mm]
$\alpha$&${+0.1360}^{+0.0419+0.0855}_{-0.0410-0.0814}$&${+0.1370}^{+0.0787+0.1621}_{-0.0758-0.1542}$&${+0.1350}^{+0.1017+0.2148}_{-0.0992-0.1993}$\\[0.4mm]
$\beta$&${+3.068}^{+0.1033+0.2129}_{-0.1026-0.2035}$&${+3.065}^{+0.0434+0.0903}_{-0.0465-0.0897}$&${+3.078}^{+0.1462+0.2953}_{-0.1556-0.3010}$\\[0.4mm]
$M$&${-19.0340}^{+0.3849+0.7605}_{-0.3907-0.7690}$&${-19.030}^{+0.4560+0.9122}_{-0.4591-0.9179}$&${-19.0310}^{+0.5241+1.0447}_{-0.5270-1.6457}$\\[0.4mm]
$dM$&${-0.120}^{+0.0299+0.2983}_{-0.2718-0.5360}$&${-0.121}^{+0.2907+0.5838}_{-0.2975-0.5906}$&${-0.117}^{+0.3290+0.6597}_{-0.3348-0.6634}$\\[0.4mm]
$\gamma_{0}$&${+0.5511}^{+0.0506+0.1010}_{-0.0375-0.0753}$&${+0.5510}^{+0.0529+0.1088}_{-0.0506-0.0985}$&${+0.5510}^{+0.0110+0.0298}_{-0.0010-0.0119}$\\[0.4mm]
$\sigma_{80}$&${+0.8180}^{+0.1400+0.2794}_{-0.1340-0.2718}$&${+0.8190}^{+0.1471+0.2987}_{-0.1504-0.3036}$&${+0.8180}^{+0.1643+0.3257}_{-0.1624-0.3213}$\\[0.4mm]
\hline\hline${\chi}^{2}_{min}$&$737.8591$&$736.8446$&$734.0572$\\[0.4mm]
\hline
\end{tabular}
\centering\begin{tabular}{| c | c | c |}
\hline
 Parameters&DR(1)&DR(2)\\
 \hline\hline
${\lambda}_{0}$&${+1.12\times10^{-4}}^{+0.6541\times10^{-4}+2.0752\times10^{-4}}_{-0.5693\times10^{-4}-1.4456\times10^{-4}}$&
${+1.12\times10^{-4}}^{+0.6541\times10^{-4}+2.0752\times10^{-4}}_{-0.5693\times10^{-4}-1.4456\times10^{-4}}$\\[0.4mm]
${\lambda}_{1}$&${+2.763\times10^{-4}}^{+0.3867\times10^{-4}+0.7578\times10^{-4}}_{-0.5252\times10^{-4}-1.8572\times10^{-4}}$&
${+2.763\times10^{-4}}^{+0.3867\times10^{-4}+0.7578\times10^{-4}}_{-0.5252\times10^{-4}-1.8572\times10^{-4}}$\\[0.4mm]
${\lambda}_{2}$&${+2.540\times10^{-5}}^{+1.0417\times10^{-5}+1.7126\times10^{-5}}_{-3.3311\times10^{-5}-3.8466\times10^{-5}}$&
${-2.586\times10^{-5}}^{+0.4648\times10^{-5}+0.9794\times10^{-5}}_{-1.1896\times10^{-5}-2.8076\times10^{-5}}$\\[0.6mm]
$w_{0}$&${-1.0364}^{+0.0644+0.1140}_{-0.0853-0.1908}$&${-1.0364}^{+0.0644+0.1140}_{-0.0853-0.1908}$\\[0.4mm]
$w_{1}$&${+2.1064}^{+0.2363+0.5842}_{-0.1213-0.1964}$&${+2.1064}^{+0.2363+0.5842}_{-0.1213-0.1964}$\\[0.4mm]
$w_{2}$&${-0.7698}^{+0.1276+0.4797}_{-0.0364-0.0717}$&${-0.7698}^{+0.1276+0.4797}_{-0.0364-0.0717}$\\[0.4mm]
$\Omega_{DM,0}$&${+0.2844}^{+0.0121+0.0385}_{-0.0061-0.0124}$&${+0.2844}^{+0.0121+0.0385}_{-0.0061-0.0124}$\\[0.4mm]
$H_{0}(kms^{-1}{Mpc}^{-1})$&${+67.1490}^{+0.8216+1.8006}_{-0.9642-1.9324}$&${+67.1490}^{+0.8216+1.8006}_{-0.9642-1.9324}$\\[0.4mm]
$\alpha$&${+0.1360}^{+0.1108+0.2341}_{-0.1198-0.2482}$&${+0.1360}^{+0.1108+0.2341}_{-0.1198-0.2482}$\\[0.4mm]
$\beta$&${+3.0780}^{+0.1968+0.3939}_{-0.1839-0.370}$&${+3.0780}^{+0.1968+0.3939}_{-0.1839-0.3700}$\\[0.4mm]
$M$&${-19.1650}^{+0.5561+1.1116}_{-0.5522-1.0996}$&${-19.1650}^{+0.5561+1.1116}_{-0.5522-1.0996}$\\[0.4mm]
$dM$&${-0.120}^{+0.3742+0.7492}_{-0.3740-0.7536}$&${-0.120}^{+0.3742+0.7492}_{-0.3740-0.7536}$\\[0.4mm]
$\gamma_{0}$&${+0.5511}^{+0.0302+0.0615}_{-0.0291-0.0571}$&${+0.5511}^{+0.0302+0.0615}_{-0.0291-0.0571}$\\[0.4mm]
$\sigma_{80}$&${+0.8190}^{+0.1706+0.3425}_{-0.1690-0.3417}$&${+0.8190}^{+0.1706+0.3425}_{-0.1690-0.3417}$\\[0.4mm]
\hline\hline${\chi}^{2}_{min}$&$731.7439$&$734.3817$\\[0.4mm]
\hline
\end{tabular}
\end{table*}
\endgroup
\begingroup
\squeezetable
\begin{table*}[!htbp]
\centering
\caption{Shows the best fitting cosmological parameters today, $f\sigma_{8,0}$, $\gamma_{0}$, ${\rm{I}}_{0}\times10^{4}$, $\omega_{0}$, $\Omega_{DM,0}$, $G_{eff,0}$, 
$H_{eff,0}$ and their errors at $1\sigma$ obtained from a combination of data.}\label{cosmological_state}
\begin{ruledtabular}
\begin{tabular}{| c | c | c | c | c | c | c | c |}
Models&$f\sigma_{8,0}$&$\gamma_{0}$&${\rm{I}}_{0}\times10^{4}$&$\omega_{0}$&$\Omega_{DM,0}$&$G_{eff,0}$&$H_{eff,0}$\\
\hline\hline
$\Lambda$CDM&${+0.4037}^{+0.0571}_{-0.0587}$&${+0.5506}^{+0.0527}_{-0.0390}$&$0.0$&$-1.0$&${+0.2810}^{+0.0185}_{-0.0138}$&$+1.0$&$+2.0$\\[0.35mm]
CPL&${+0.4049}^{+0.0580}_{-0.0585}$&${+0.5505}^{+0.0551}_{-0.0527}$&$0.0$&${-1.0323}^{+0.0489}_{-0.0605}$&${+0.2819}^{+0.0171}_{-0.0094}$&$+1.0$&$+2.0$\\[0.35mm]
XCPL&${+0.4063}^{+0.1035}_{-0.0965}$&${+0.5506}^{+0.0114}_{-0.0069}$&${+6.8730}^{+2.2096}_{-1.9077}$&${-1.0271}^{+0.0563}_{-0.0610}$&
${+0.2840}^{+0.0308}_{-0.0290}$&${+1.000146}^{+3.07\times{10}^{-4}}_{-3.9\times{10}^{-4}}$&${+1.999902}^{+3.5\times{10}^{-5}}_{-4.5\times{10}^{-5}}$\\[0.35mm]
DR(1)&${+0.4070}^{+0.0781}_{-0.0758}$&${+0.5507}^{+0.0313}_{-0.0302}$&${+0.8660}^{+0.5543}_{-0.2362}$&${-1.0364}^{+0.0644}_{-0.0853}$&
${+0.2844}^{+0.0121}_{-0.0061}$&${+0.999892}^{+2.0\times{10}^{-5}}_{-1.9\times{10}^{-5}}$&${+1.999988}^{+3.0\times{10}^{-6}}_{-9.0\times{10}^{-6}}$ \\[0.35mm]
DR(2)&${+0.4070}^{+0.0781}_{-0.0758}$&${+0.5507}^{+0.0313}_{-0.0302}$&${+1.3786}^{+0.6070}_{-0.4503}$&${-1.0364}^{+0.0644}_{-0.0853}$&
${+0.2844}^{+0.0121}_{-0.0061}$&${+0.999883}^{+2.3\times{10}^{-5}}_{-1.8\times{10}^{-5}}$&${+1.999980}^{+7.0\times{10}^{-6}}_{-9.0\times{10}^{-6}}$ \\[0.35mm]
\end{tabular}
\end{ruledtabular}
\end{table*}
\endgroup
\begingroup
\squeezetable
\begin{table*}[!htbp]
\centering
\caption{Shows the $z_{cross}$ points and the values of $f\sigma_{8}$, $\gamma$, ${\bar{\rm{I}}_{\rm Q}}$, $\omega$, $\Omega_{DM}$, $G_{eff}$, 
$H_{eff}$ evaluated at $z_{cross}$.}\label{crossing_state}
\begin{ruledtabular}
\begin{tabular}{| c | c | c | c | c | c | c | c |}
Models&$z_{cross}$&$f\sigma_{8}(z_{cross})$&$\gamma(z_{cross})$&$\omega(z_{cross})$&$\Omega_{DM}(z_{cross})$&$G_{eff}(z_{cross})$&$H_{eff}(z_{cross})$\\
\hline\hline
XCPL&$+0.8886$&$+0.4348$&$+0.4734$&$-0.9824$&$+0.6571$&$+1.000475$&$+2.0$\\[0.35mm]
DR1&$-0.3342$&$---$&$---$&$-1.8518$&$+0.1281$&$+0.999946$&$+2.0$  \\[0.35mm]
DR2&$-0.4593$&$---$&$---$&$-2.1799$&$+0.080$&$+0.999940$&$+2.0$  \\[0.35mm]
DR2&$+5.8073$&$+0.1949$&$-14.7150$&$-3.0065$&$+0.8357$&$+1.000736$&$+2.0$ \\[0.35mm]
\end{tabular}
\end{ruledtabular}
\end{table*}
\endgroup
\begin{figure*}[!htb]
 \centering\includegraphics[width=17cm,height=9cm]{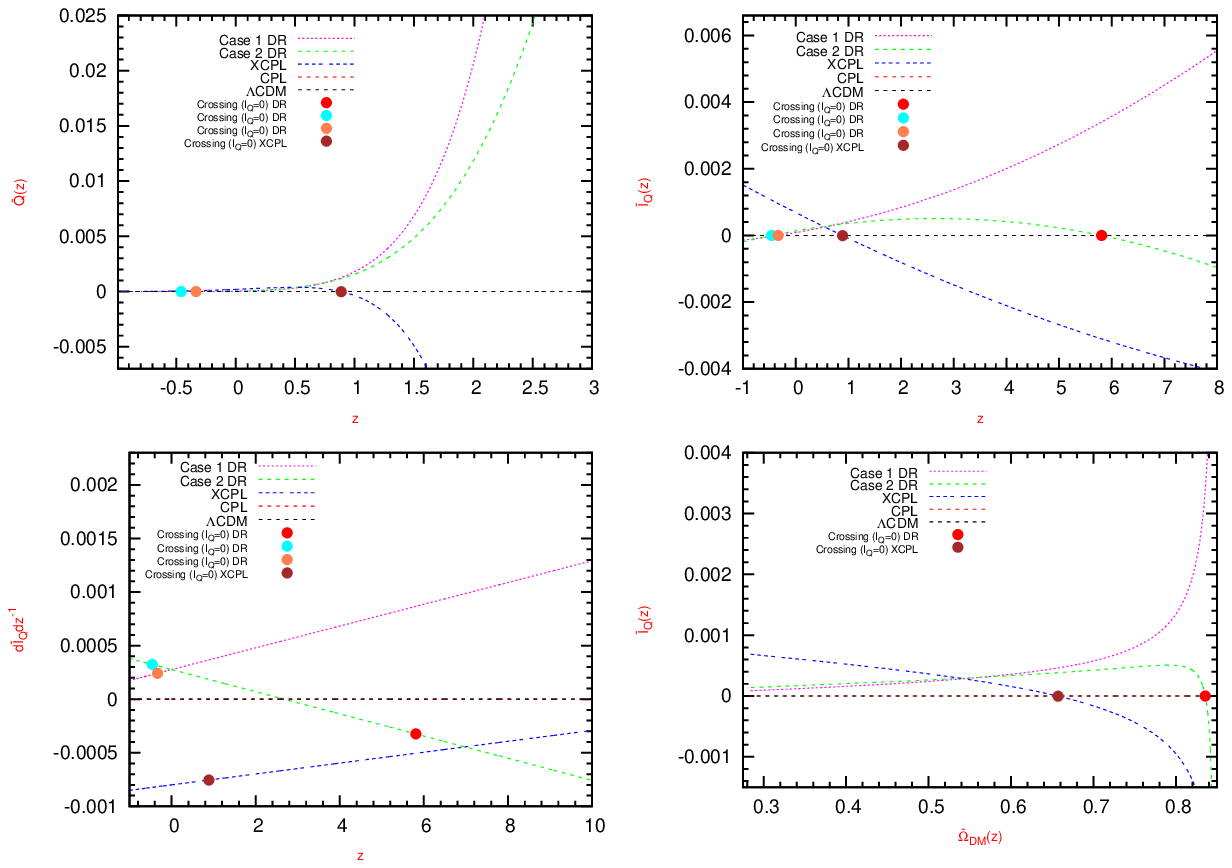}\\
 \caption{(color online) The upper panels display the best reconstructed $\bar{Q}(z)$ and ${\bar{\rm I}}_{\rm Q}$ along $z$. Similarly, the lower panels show the reconstructed evolution 
 of $\mathrm{d}{\bar{\rm I}}_{\rm Q}{\mathrm{d}z}^{-1}$ in function of $z$ and the effect of ${\bar{\rm{I}}}_{\rm Q}$ on ${\bar{\Omega}}_{DM}$, respectively.}
 \label{Behaviours_IQ}
\end{figure*}
\begin{figure*}[!htb]
 \centering \includegraphics[width=17cm,height=9cm]{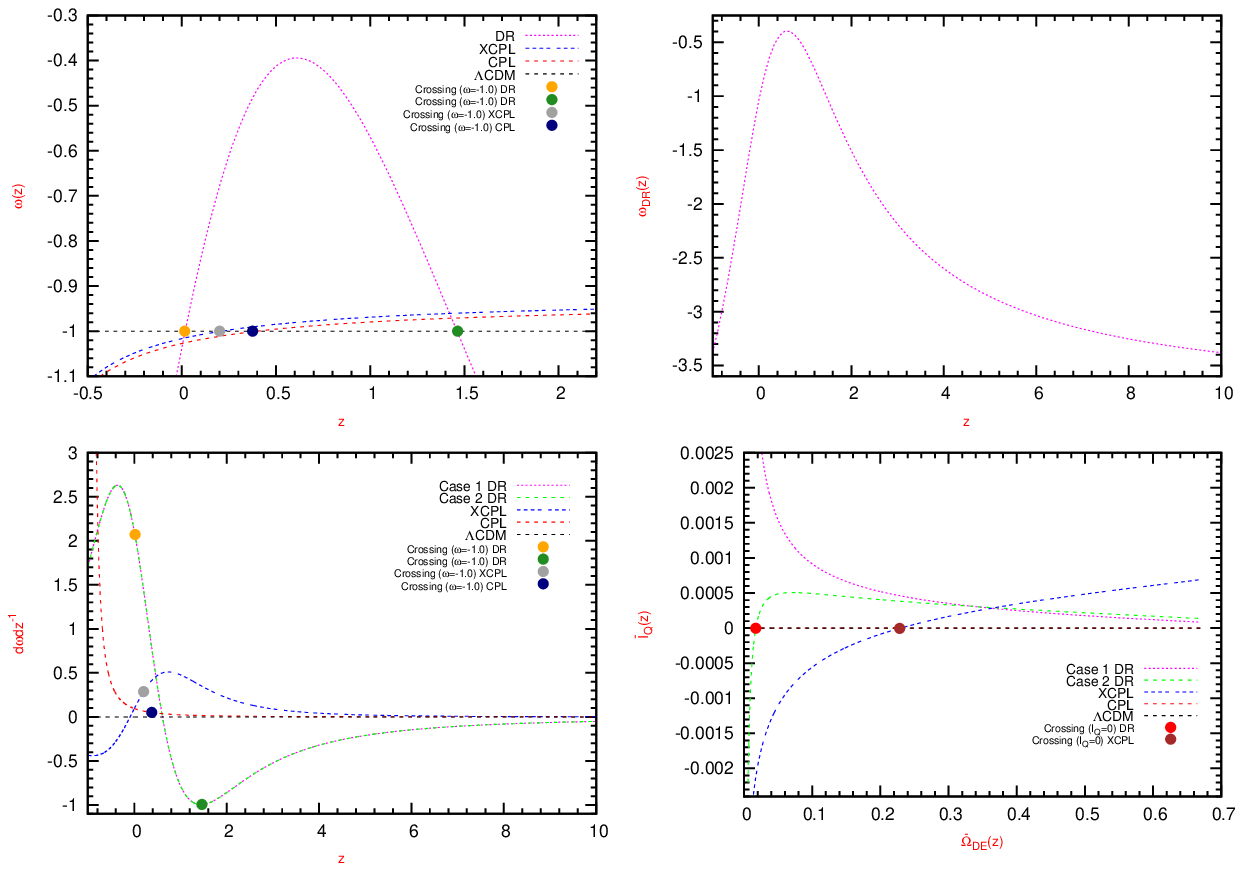}\\
 \caption{(color online) The upper panels display the best reconstructed $\omega(z)$ and $\omega_{DR}(z)$ along $z$. Similarly, the lower panels show the reconstructed 
 evolution  of $\mathrm{d}{\omega}{\mathrm{d}z}^{-1}$ in function of $z$ and the effect of ${\bar{\rm{I}}}_{\rm Q}$ on ${\bar{\Omega}}_{DE}$, respectively.}
 \label{Behaviours_w}
\end{figure*}
\begin{figure*}[!htb]
 \centering\includegraphics[width=17cm,height=9cm]{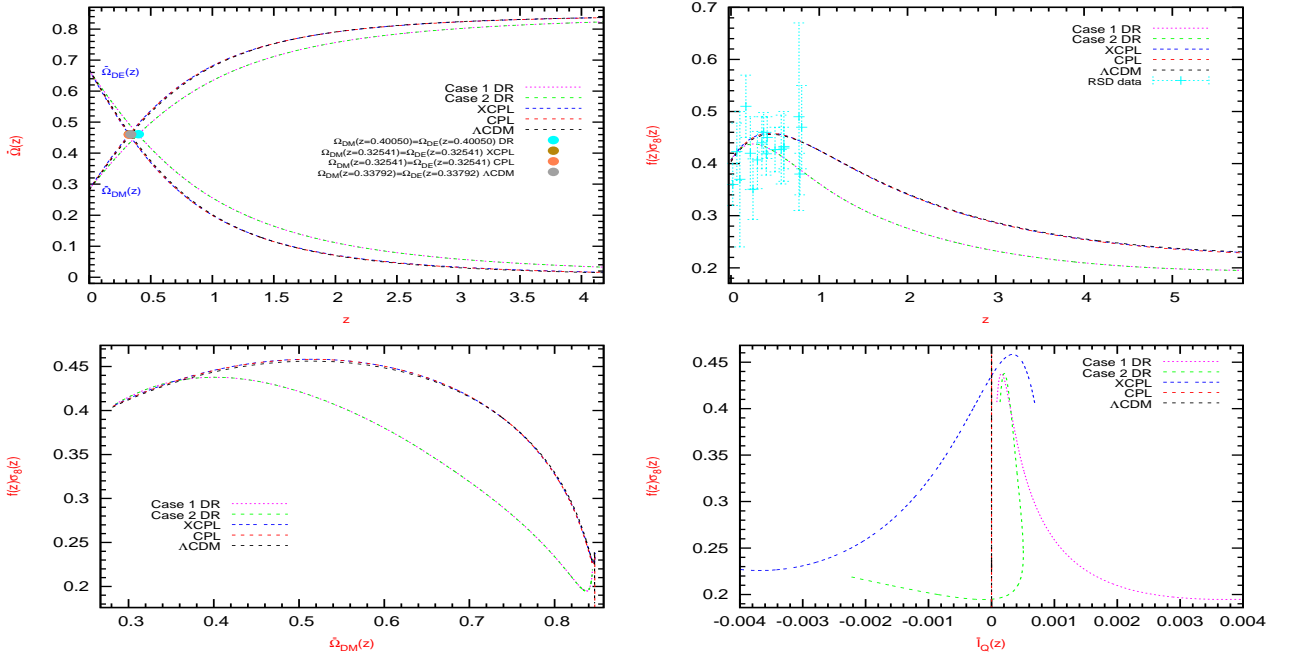}\\
 \caption{(color online) The left above panel shows the evolution of ${\bar{\Omega}}$ along $z$, whereas the right panel displays the evolution of $f\sigma_{8}$ as function 
 of $z$. The left and right below panels depict the evolution of $f\sigma_{8}$ as function of ${\bar{\Omega}}_{DM}$ and ${\bar{\rm{I}}}_{\rm Q}$, respectively.}
 \label{Effects1}
\end{figure*}
\begin{figure*}[!htb]
 \centering\includegraphics[width=17cm,height=9cm]{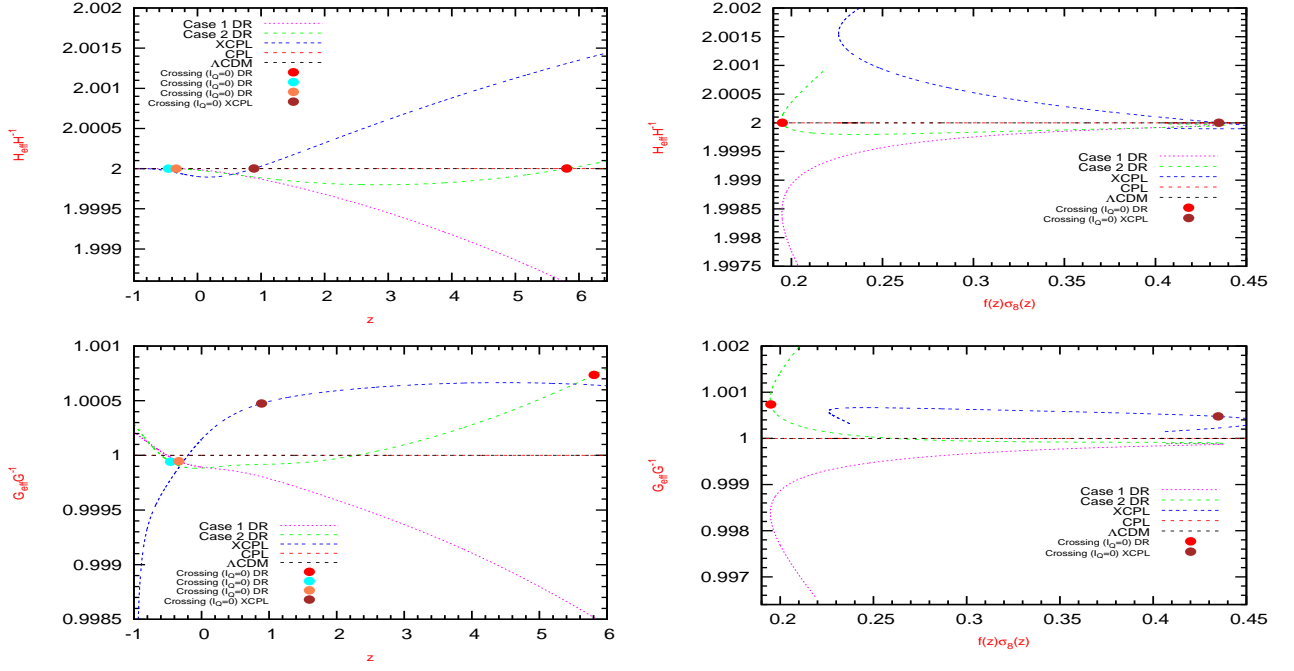}\\
 \caption{(color online) The left above panel shows the evolution of $H_{eff}H^{-1}$ along $z$, whereas the right above panel displays the effect the frictional force on the 
 evolution of $f\sigma_{8}$. By contrast, the left and right below panels depict the same but for $G_{eff}G^{-1}$.}\label{Effects2}
\end{figure*}
\begin{figure*}[!htb]
 \centering\includegraphics[width=17cm,height=9cm]{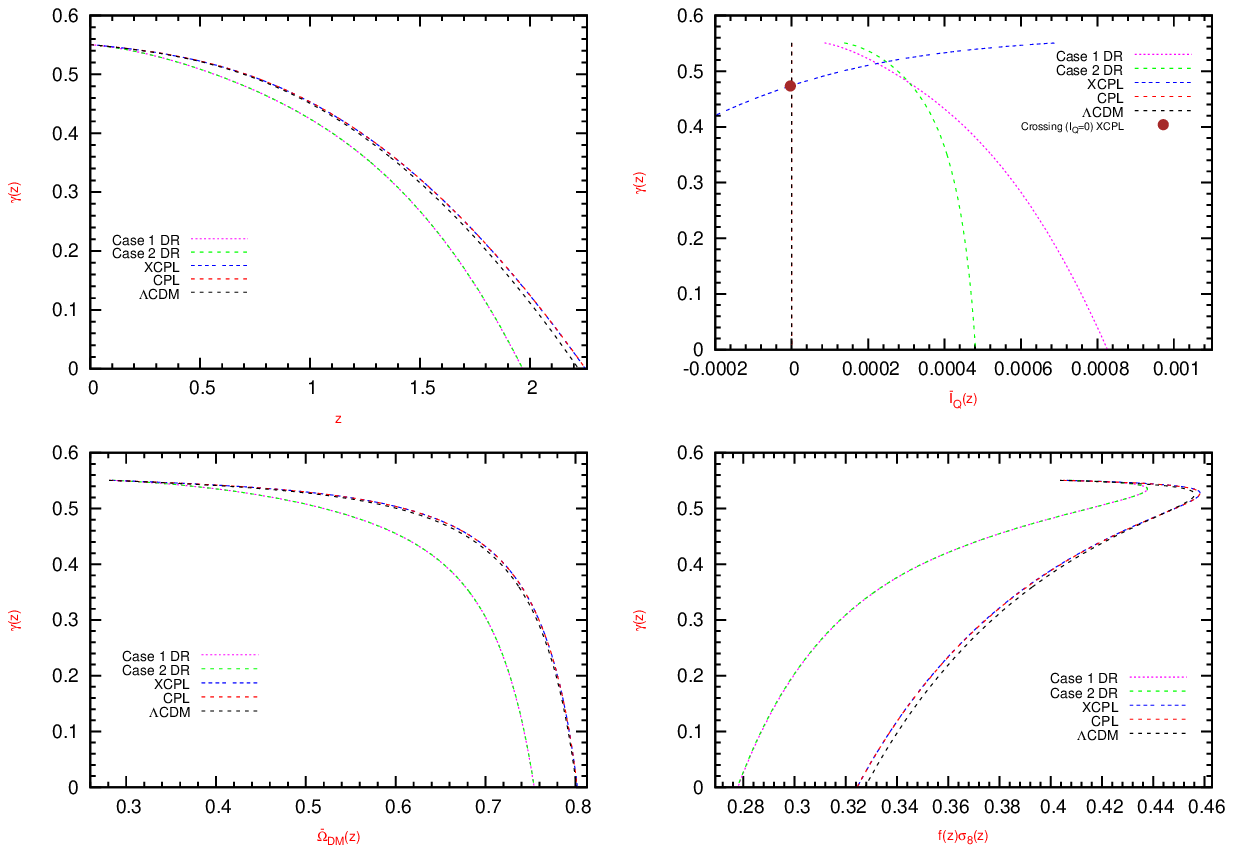}\\
 \caption{(color online) The left above panel shows the evolution of $\gamma$ along $z$, whereas the right above panel displays the effect of ${\bar{\rm{I}}}_{\rm Q}$ on the 
 evolution of $\gamma$. In addition, the left below panel depicts the effect of ${\bar{\Omega}}_{DM}$ on $\gamma$, and the effect of $\gamma$ on the cosmic structure 
 formation is shown in the right below panel.}\label{Effects3}
\end{figure*}
\section{Results}
We constructed a code to calculate numerically the theoretical evolutions of $\delta$ and $f$, respectively, and therefore, the values of $f\sigma_{8}$, setting functional 
forms on ${\bar{\rm I}}_{\rm Q}$ and $\omega$ such that they can be easily implemented in each of our models. Then, via a Markov Chain Monte Carlo (MCMC) 
analysis, we can perform a global fitting in each of them (listed in Table \ref{Bestfits}), by using a combined statistical analysis of cosmic observations such as 
JLA data, the RSD data, the BAO data, the CMB given by the Planck $2015$ data and the H data; from which, we could reduce the uncertainty and put tighter 
constraints on the values of the cosmological parameters. Table \ref{Priors} describes the priors used in this work. For each of the models, the one-dimension probability 
contours, the best fitting parameters and their errors (at $1\sigma$ and $2\sigma$) are shown in Fig. \ref{Contours}.\\
The values of the functions $f\sigma_{8}$, $\gamma$, ${\bar{\rm I}}_{\rm Q}$, $\omega$, ${\bar{\Omega}}_{DM}$, $G_{eff}$ and $H_{eff}$ evaluated in $z=0$ (today) are denoted 
as $f\sigma_{8,0}$, $\gamma_{0}$, ${\rm I}_{0}$, $\omega_{0}$, ${\bar{\Omega}}_{DM,0}$, $G_{eff,0}$ and $H_{eff,0}$, respectively (see Table \ref{cosmological_state}.\\
In the following Figs. the constraints at $1\sigma$ and $2\sigma$ on ${\bar{\Omega}}_{DM}$, ${\bar{\Omega}}_{DE}$, ${\bar{\rm I}}_{\rm Q}$, $\omega$, $f\sigma_{8}$, $H_{eff}H^{-1}$, 
$G_{eff}G^{-1}$ and $\gamma$ have been omitted to obtain a better visualization of the results.\\
Let us now see Fig. \ref{Behaviours_IQ}, within the coupled models have considered that ${\rm I}_{+}$ denotes an energy transfer from $DE$ to $DM$; on the contrary, ${\rm I}_{-}$ 
denotes an energy transfer from $DM$ to $DE$. In this regard, within the coupled models have found a change from ${\rm I}_{+}$ to ${\rm I}_{-}$ and vice versa. 
A change of sign on the best reconstructed ${\bar{\rm I}}_{\rm Q}$ is linked to the crossing of the non-coupling line, ${\bar{\rm I}}_{\rm Q}(z)=0$. Table \ref{crossing_state} 
shows the $z=z_{crossing}$ points that satisfy the condition ${\bar{\rm I}}_{\rm Q}(z)=0$, which were already predicted by Eq. (\ref{ZcrossIq}). Moreover, the left below panel 
in Fig. \ref{Behaviours_IQ}, confirms the statement given by Eq. (\ref{dIqdznull}). We also verify that if the $z$ points satisfy the relation ${{\mathrm{d}{\bar{\rm I}}_{\rm Q}}{\mathrm{d}z}^{-1}}|_{z}=0$, 
then they will be different in comparison with the $z_{cross}$ points.
According to Table \ref{cosmological_state} and the upper panels in Fig. \ref{Behaviours_IQ}, note that a non-negligible value of ${\bar{\rm I}}_{0}$ at $1\sigma$ error is found in the coupled 
models, and whose order of magnitude is in agreement with the results obtained in \cite{abramo1,Cai-Su,abramo2,cao2011,LiZhang2011,Cueva-Nucamendi2012}.
However, due to the two minimums obtained in the DR model (see Table \ref{Bestfits}), two different cases ($1$ and $2$) to reconstruct ${\rm I}_{\rm Q}$ are worked here. 
From Table \ref{Bestfits} we focus on the case $2$, which is in disagreement with the result obtained in Eq. (\ref{dIqdznull7}); in this way, the observational data are the 
fundamental tool to fix the constraints on the cosmological parameters, testing and choosing the possible theoretical models to be worked.
On the other hand, from the results presented in Fig. \ref{Behaviours_w}, we note that in the left and right above panels the universe evolves from the quintessence regime 
$\omega > -1$ to the phantom regime $\omega < -1$, and in particular, crosses the phantom divide line $\omega(z_{phantom})=-1$ \cite{Nesseris2007}. In the DR model, this 
crossing feature is more favored with two phantom crossing points in $z=z_{phantom\,1}=0.0155$ and $z=z_{phantom\,2}=1.4643$, respectively, instead, the XCPL model shows only one 
phantom crossing point in $z=z_{phantom\,3}=0.2003$. Likewise, the CPL model also depicts one phantom crossing point in $z=z_{phantom\,4}=0.3755$.
From these above panels in Fig. \ref{Behaviours_w}, we also see that in the XCPL model the evolution of $\omega$ is similar to that in the CPL model; in contrast, the parameter 
$\omega$ defined in the DR model, starts to evolve from the value $\omega=5\omega_{2}$ during the matter era and reaches the value $\omega=\omega_{0}$ in the present time. Likewise, a finite value 
$\omega(z=-1)=(5/3)\omega_{2}+(2/3)[\omega_{0}-\omega_{1}]$ is obtained in the future. We stress that there is a significant difference for the evolution of $\omega$ 
in the XCPL and DR models, and depend on the epoch at which they are compared. From the right above panel in Fig. \ref{Behaviours_w}, we find that in the DR 
model when $0.6133\leq z \leq10$, the amplitude of $\omega$ grows from $-3.3799$ to $-0.3942$. By contrast, when $-1.0< z\leq 0.6133$, the amplitude of $\omega$ decreases 
from $-0.3942$ to $-3.3781$, whereas in the XCPL model for $-1<z \leq 10$, the amplitude of $\omega$ decreases more rapidly than that in the DR model. Indeed, these characteristics 
are a consequence of the reconstructed EoS parameters in the CPL, XCPL and DR models, respectively. In addition, in the DR model for the region $0\leq z \leq1.5$, $\omega$ deviates 
significantly from $\omega=-1$, with a pronounced peak at around $z=0.6133$ and with an average value of $\omega=-0.3942$. This behavior is opposite with the evolution of 
$\omega$ in \cite{Lu2009-Neveu2013}.\\
From the left below panel in Fig. \ref{Behaviours_w}, we also verify that if the $z$ points satisfy the relation $\omega(z_{phantom})=-1$, then 
one finds the following condition ${\mathrm{d}}{\omega}{\mathrm{d}z}^{-1}|_{z_{phantom}}\neq 0$.\\
The right below panels in Figs. \ref{Behaviours_IQ} and \ref{Behaviours_w}, show that ${\bar{\rm I}}_{\rm Q}$ could take positive or negative values during its evolution 
from $-0.0025$ to $+0.0025$. Therefore, the values for ${\bar{\Omega}}_{DM}$ moves from $0.85$ to $0.284$, and the values for ${\bar{\Omega}}_{DE}$ moves from $0$ to $0.67$, 
respectively. These final values for ${\bar{\Omega}}_{DM}$ and ${\bar{\Omega}}_{DE}$ are indicated in Table \ref{Bestfits}.\\

From the upper panels in Figs. \ref{Behaviours_IQ} and \ref{Behaviours_w}, and from left above panel in Fig. \ref{Effects1}, we focus on the DR model at 
$0.6133\leq z \leq10$. Here, $\omega$ grows and ${\bar{\rm I}}_{\rm Q}$ could take positive, negative ad null values, and therefore, they will force to the fact that the 
concentration of $\Omega_{DE}$\,(${\bar{\Omega}}_{DM}$) to grow (decrease) more rapidly than those in the $\Lambda$CDM, CPL and XCPL models, respectively. For $-1<z<0.6133$, 
$\omega$ decreases and ${\bar{\rm I}}_{\rm Q}$ could take positive, negative and null values. Thus, they will induce to the fact that the values obtained for 
${\bar{\Omega}}_{DE}\,({\bar{\Omega}}_{DM})$ in $z=0$ are closer to those values measured today, with $DE$ is dominant.\\
Considering the right above panel in Fig. \ref{Behaviours_IQ}, the right below panels in Figs. \ref{Behaviours_IQ} and \ref{Behaviours_w}, and the left above panel in 
Fig. \ref{Effects1}, we see that for $z\geq 0.3254$ the value of the amplitude of ${\bar{\Omega}}_{DM}$ (${\bar{\Omega}}_{DE}$) in the DR model is slightly modified by 
the values of ${\bar{\rm I}}_{\rm Q}$ (${\rm I}_{+}$ and ${\rm I}_{-}$) relative to the other model, it means that, ${\bar{\rm I}}_{\rm Q}$ changes from ${\rm I}_{+}$ to ${\rm I}_{-}$, and vice 
versa. In this model the amplitude of ${\bar{\Omega}}_{DM}$\,(${\bar{\Omega}}_{DE}$) is suppressed (amplified) in comparison with those found in the other models. This result 
coincides with that found in \cite{cabral2009}. Here, we also confirm that the coincidence problem is alleviated in these coupled models, but they may not solve it.\\
From Table \ref{cosmological_state} and from the below panels in Fig. \ref{Effects2}, note that the values of $G_{eff}G^{-1}$ deviate significantly from unity in all $z$. It is 
in agreement with the resulted found in \cite{Amendola2004,Das2006}. Accordingly, $G_{eff}G^{-1}$ are growing or decreasing functions, and could cross the value 
$G_{eff}G^{-1}=1$ at less one time or twice. Furthermore, at $z=0$, the values of $G_{eff}G^{-1}$ can be roughly larger than $1$ (XCPL model) or smaller than $1$ (DR model). 
These observations show the effects of the reconstructions of ${\bar{\rm I}}_{\rm Q}$ and $\omega$ on the evolution of $G_{eff}G^{-1}$ in the linear regime.\\
Similarly, considering Table \ref{cosmological_state}, the right above and below panels in Fig. \ref{Effects1} and the upper panels in Fig. \ref{Effects2}, find the effect 
of ${\bar{\Omega}}_{DM}$ and ${\bar{\rm I}}_{\rm Q}$ on the evolution of $H_{eff}H^{-1}$ function. At around $z=0$, the values of $H_{eff}H^{-1}$ in the 
coupled models are roughly different among them. Moreover, for $-1\leq z \leq6$ the best fitting of $H_{eff}H^{-1}$ in the XCPL model deviates significantly with 
respect to that obtained in the DR  (cases $1$ and $2$) model. This is a consequence of the higher quantity of ${\bar{\Omega}}_{DM}$ concentrated and of a lesser magnitude 
of $H_{eff}H^{-1}$ in the XCPL model. Therefore, in the region $0\leq z \leq6$, the amplitude of $f\sigma_{8}$ is higher and more pronounced in the XCPL with respect 
to that in the DR model. At the end of the matter era (at $z\approx 0.4$) the value of $H_{eff}H^{-1}$ in the XCPL model decreases to be approximately smaller than those in 
the DR and uncoupled models, respectively. In the regime $-0.5 \leq z<0.4$, the values of $H_{eff}H^{-1}$ in the DR model become closer to $2$, and thus higher than 
those found in the XCPL model, reducing the cosmic structure formation (see right above panel in Fig. \ref{Effects1}). 
Likewise, according to Eqs. (\ref{Geff}) and (\ref{Heff}) an increase on the magnitudes of ${\bar{\rm I}}_{\rm Q}$ and ${\bar{\Omega}}_{DE}$ tend to amplify the 
gravitational strength ($G_{eff}G^{-1}$), but they reduce the magnitude of the frictional force. In fact, an increase in $G_{eff}$ would enhance the growth of structure even 
at later times.\\
In addition, the right above and right below panels in Fig. \ref{Effects2}, show that $H_{eff}H^{-1}$ ($G_{eff}G^{-1}$) during its evolution could take values from 
$1.9975$ to $+2.0020$ (from $0.9970$ to $+1.0020$), and therefore the value of $f\sigma_{8}$ could move from $0.19$ to $0.4583$. This final value for $f\sigma_{8}$ is shown 
in the right upper panel of Fig. \ref{Effects1}, and indicated in Table \ref{cosmological_state}.\\ 
From Figs. \ref{Effects1} and \ref{Effects2}, we consider the evolution of ${\bar{\rm I}}_{\rm Q}$ and ${\bar{\Omega}}_{DE}$, finding explicitly that in the XCPL model 
the functions $G_{eff}G^{-1}$ follows a different behavior from that predicted in the DR model. For this reason, the deviations of $G_{eff}G^{-1}$ from standard gravity are 
significant. It also explains why $f\sigma_{8}$ is larger in the XCPL model (and uncoupled models) than that in the DR model; especially, when $0\leq z \leq 6$. More explicitly, for $0<z<1$, 
the values of $G_{eff}G^{-1}$ are close to $1$ in the DR model and larger than $1$ in the XCPL model, implying the existence of a non-standard gravity. Therefore, in the 
coupled models the evolution of ${\bar{\rm I}}_{\rm Q}$, ${\bar\Omega}_{DE}$, ${\bar\Omega}_{DM}$ and $\omega$ are different such that their effects cannot be 
ruled out. The modifications to gravity enhance the structure formation at late times in the XCPL models, but suppresses it in the DR model, when $0\leq z\leq 6$. In general, 
the deviation of $G_{eff}G^{-1}$ from unity starts at early times ($z \geq 1$) in the coupled models. This indicates that the magnitude of ${\bar{\rm I}}_{\rm Q}$ is very 
large there. Thus, for $z>1$, the substantial difference in the values of $G_{eff}G^{-1}$ is more pronounced. In other words, in the XCPL model the matter density is 
much higher than that in the DR model, and therefore affecting more the magnitude of $f\sigma_{8}$ in the XCPL model than that in the DR model. This explains why the results are 
very different in this regime, and the differences from uncoupled models are induced mainly by the effective Hubble friction term, $H_{eff}H^{-1}$ (which acts as a frictional 
force that slows down the linear structure growth).\\
The left above panel in Fig. \ref{Effects3}, depicts the evolution of $\gamma$ along $z$ for the coupled and uncoupled models. Likewise, the right upper and right below 
panels in Fig. \ref{Behaviours_IQ}, show that ${\bar{\rm I}}_{\rm Q}$ could take positive or negative values during its evolution from $z=8.0$ to $z=-1$, and the values for 
${\bar{\Omega}}_{DM}$ moves from $0.8$ to $0.284$. From here, and using the right upper and left below panels in Fig. \ref{Effects3} note that the amplitude for $\gamma$ is 
progressively increased to become approximately $\gamma=0.56$. Additionally, from the right below panel of this Figure and considering the coupled models, note that the 
values for growth of cosmic structure are very different in the past, and hence the corresponding values for $\gamma$ are very closed to zero. If the values for $f\sigma_{8}$ 
are progressively increased, then the values for $\gamma$ also increase, and become much more stable, when $\gamma \approx \gamma_{0}$. In Table \ref{cosmological_state} 
show the values of $\gamma_{0}$ for each of the models studied.\\
Let us analyze the right upper and left below panels in Fig. \ref{Effects3}. From here, we find that the magnitude of ${\bar{\rm I}}_{\rm Q}$ has imprinted new physical effects 
on the evolution of the parameter, $\gamma$. In the DR model the amplitude of $\gamma$ is progressively reduced in the region $0\leq z \leq 2.3$, with respect to those 
found in the uncoupled models. Therefore, this shows that the magnitude of ${\bar{\rm I}}_{\rm Q}$ is strongly related with the magnitudes of ${\bar\Omega}_{DM}$, 
$f\sigma_{8}$ and $\gamma$, respectively.\\
We now compare our results with those obtained by other researchers. In \cite{Bueno2011}, the authors parameterized $\gamma$ in terms of the Legendre polynomials, and 
compared it with those obtained from other cosmological models. Here power spectrum data and weak lensing power spectrum data were used. Our results obtained for $\gamma$ 
are very closed to that obtained in the $F(R)$ model, at $1\sigma$ error. Furthermore, in \cite{Alcaniz2013}, the authors provided a convenient analytic formula for 
$f\sigma_{8}$, which was applied to different $DE$ models. They used RSD data to place observational constraints. The results obtained by them on $f\sigma_{8}$ are consistent 
at $1\sigma$ error with our results. Likewise, Pouri et al. in \cite{Pouri2014} used the clustering properties of Luminous Red Galaxies (LRGs) and the growth rate data to 
constrain $\gamma$. The results found by them on $\gamma$ and $f\sigma_{8}$ are compatible with our results, at $1\sigma$ error. Similarly, Yang and Xu in \cite{Yang2014}, 
studied a model composed by the cosmological constant, with a nonzero $DM$ EoS parameter. The result obtained by on $f\sigma_{8}$ is consistent with our result at 
$1\sigma$ error. Also, the authors in \cite{Mehrabi2015}, studied the impact of $DE$ clustering on $\gamma$. They used two different EoS parameters, and found a fitting 
evolution curve for $f\sigma_{8}$, which at $1\sigma$ error is acceptable with our result.
\section{Conclusions}\label{SectionConclusions}
Now we summarize our main results:\\
$\bullet$ An analysis combined of data was performed to break the degeneracy among the different cosmological parameters of our models, obtaining constraints more 
stringent on them. In particular, for the XCPL and DR models, the allowed regions for their parameters are significantly reduced by the inclusion of the CMB and RSD data 
when are compared with studies of models without these data \cite{Cueva-Nucamendi2012}. This implies that higher redshift and dynamical probes may be able to discriminate 
between these models.\\
$\bullet$ In the DR model, a novel reconstruction for $\omega$ is proposed here, and whose best fitted value is closed to $-1$. Moreover, it has the property of avoiding 
divergences in a distant future $z\rightarrow -1$. This result is consistent with the value predicted by the $\Lambda$CDM model at $1\sigma$ error. Likewise, within this 
coupled scenario, a finite value for $\omega$ is obtained from the past to the future; namely, the following asymptotic values are found: $\omega(z)=5\omega_{2}$ for $z\gg1$, 
$\omega(z)\approx \omega_{0}$ for $z\ll 1$ and $\omega(z)\approx(5/3)\omega_{2}+(2/3)[\omega_{0}-\omega_{1}]$ for $z\rightarrow -1$ (see right above panel in 
Fig. \ref{Behaviours_w}). Therefore, a possible physical description performed by the DR model on the dynamical evolution of $DE$ should be used to explore its 
properties.\\
$\bullet$ In the coupled models the values of the amplitudes of ${\bar{\Omega}}_{DE}$ (see left upper panel in Fig.\,\ref{Effects1}) are slightly modified by the 
reconstructions of ${\bar{\rm I}}_{\rm Q}$ and $\omega$ when they are compared with those in the uncoupled models. Nevertheless, they are definitely positive. 
This requirement implies that $\omega$ must be always negative in all the cosmic stages of the universe (see upper panels in Fig. \ref{Behaviours_w}).\\
$\bullet$ If ${\bar{\rm I}}_{\rm Q}$ takes the values ${\rm I}_{+}$ and ${\rm I}_{-}$ (see right upper panel in Fig. \ref{Behaviours_IQ}), then the amplitudes of 
${\bar{\Omega}}_{DM}$ (${\bar{\Omega}}_{DE}$) (see left upper panel in Fig. \ref{Effects1}) for the two cases in the DR model are smaller (larger) in the past than 
their corresponding ${\bar{\Omega}}_{DM}$ (${\bar{\Omega}}_{DE}$) in the XCPL and uncoupled models. Likewise, we also found in the DR model that the values of the 
amplitudes of ${\bar{\Omega}}_{DE}$) are significantly affected by the values of both ${\rm I}_{\rm Q}$ and $\omega$. Naturally, a smaller proportion of $DM$ leads to a 
lesser cosmic structure formation. Therefore, the magnitude of $f\sigma_{8}$ in the DR model is suppressed in comparison with those found in the XCPL and uncoupled models 
(see right upper panel in Fig. \ref{Effects1}).\\
$\bullet$ For different redshifts, we note that in the coupled models the evolution of $H_{eff}H^{-1}$ and $G_{eff}G^{-1}$ (see Fig. \ref{Effects2}) follow different behaviors from those 
found in the uncoupled models. Therefore, they represent a deviation from the evolution predicted by the uncoupled models. Consequently, the DR model predicts an enhancement 
(suppression) on the amplitude of ${\bar{\Omega}}_{DE}$ (${\bar{\Omega}}_{DM}$) with respect to that found in the XCPL model (see left upper panel in Fig.\,\ref{Effects1}). 
These effects are significantly sensible to the reconstructions of ${\bar{\rm I}}_{\rm Q}$ and $\omega$, respectively, and decrease when $z$ tends to zero.\\
$\bullet$ In the coupled $DE$ models, the decisive role in modifying the cosmic structure formation relative to the uncoupled models is determined mainly by the evolution 
of ${\bar{\rm I}}_{\rm Q}$, $\omega$ and ${\bar{\Omega}}_{DM}$, respectively. For $z=0$, the values of $f\sigma_{8}$ are very closed to each other.\\
$\bullet$ Currently, an enhancement on the amplitude of $f\sigma_{8}$ is the situation revealed in XCPL model when it is compared with that in the DR model, and therefore 
these scenarios should be considered to study new physical properties of the universe (see right upper panel in Fig.\,\ref{Effects1}).\\
$\bullet$ The behaviors qualitatively presented here show that the plot for $\gamma$ has more possibility in discriminating the different coupled $DE$ models, 
and therefore $\gamma$ could be used to distinguish them (see left upper panel in Fig. \ref{Effects3}).\\ 
\appendix
\section*{Apendixes}
\section{Integrals $I_n(z)$ and $\tilde{I_{n}}(\tilde{x})$}\label{integralIn}
\begin{widetext}
\begin{eqnarray}
I_{0}(z)&=&\frac{2}{z_{max}}\biggl[\ln\biggl(1+z\biggr)\biggr]\;,\\
I_{1}(z)&=&\frac{2}{z_{max}}\biggl[\frac{2z}{z_{max}}-\frac{(2+z_{max})}{z_{max}}\ln\biggl(1+z\biggr)\biggr]\;,\\
I_{2}(z)&=&\frac{2}{z_{max}}\biggl[\frac{4 z}{z_{max}}\left(\frac{z}{z_{max}}-\frac{2}{z_{max}}-2\right)+
\left(1+\frac{6.8284}{z_{max}}\right)\left(1+\frac{1.1716}{z_{max}}\right)\ln\biggl(1+z\biggr)\biggr]\;,\\
\tilde{I_{0}}(\tilde{x})&=&\frac{2}{z_{max}}\biggl[\ln{\biggl(1+0.5\;z_{max}(1+\tilde{x})\biggr)}\biggr]\;,\\
\tilde{I_{1}}(\tilde{x})&=&{\frac{2}{z_{max}}}\biggl[\biggl(1+\tilde{x}\biggr)-\frac{(2+z_{max})}{z_{max}}\ln{\biggl(1+0.5\;z_{max}(1+\tilde{x})\biggr)}\biggr]\;,\\
\tilde{I_{2}}(\tilde{x})&=&{\frac{2}{z_{max}}}\biggl[\biggl(1+\tilde{x}\biggr)\biggl(\tilde{x}-\frac{4}{z_{max}}-3\biggr)+\biggl(1+\frac{6.8284}{z_{max}}\biggr)
\biggl(1+\frac{1.1716}{z_{max}}\biggr)\ln{\biggl(1+0.5\;z_{max}(1+\tilde{x})\biggr)}\biggr]\;.
\end{eqnarray}
\end{widetext}
\section{Quantities $A_n$ and $J_{n}$}\label{integralAnJn}
\begin{widetext}
\begin{eqnarray}
A_{0}&=&\frac{8}{(2+z_{max})^{2}}\biggl[\omega_{0}-\omega_{1}+\omega_{2}+8\omega_{2}(z_{max}^{-2}+z_{max}^{-1})-2\omega_{1}z_{max}^{-1}\biggr]\;,\\
A_{1}&=&\frac{4}{(2+z_{max})^{2}}\biggl[\omega_{1}-\omega_{0}+5\omega_{2}+16\omega_{2}(z_{max}^{-2}+z_{max}^{-1})+2\omega_{1}z_{max}^{-1}\biggr]\;,\\
A_{2}&=&\frac{4}{(2+z_{max})^{2}}\biggl[\omega_{0}+2\omega_{1}-5\omega_{2}-16\omega_{2}(z_{max}^{-2}+z_{max}^{-1})+4\omega_{1}z_{max}^{-1}\biggr]\;,\\
J_{0}&=&\frac{-4}{3(2+z_{max})^{2}}\biggl[\lambda_{0}-\lambda_{1}-\lambda_{2}+2\lambda_{2}(1+2z_{max}^{-1})^{2}-2\lambda_{1}z_{max}^{-1}\biggr]\;,\\
J_{1}&=&\frac{-2}{3(2+z_{max})^{2}}\biggl[-\lambda_{0}+\lambda_{1}+\lambda_{2}+4\lambda_{2}(1+2z_{max}^{-1})^{2}+2\lambda_{1}z_{max}^{-1}\biggr]\;,\\
J_{2}&=&\frac{-2\sqrt2}{3(2+z_{max})^{2}}\biggl[\lambda_{0}+2\lambda_{1}-\lambda_{2}-4\lambda_{2}(1+2z_{max}^{-1})^{2}+4\lambda_{1}z_{max}^{-1}\biggr]\;.
\end{eqnarray}
\end{widetext}
\begin{acknowledgments}
The author is grateful to Prof. F. Astorga for his academic support and fruitful discussions in the early stages of this research, and also thank Prof. O. Sarbach for useful 
comments. This work was in beginning supported by the IFM-UMSNH.
\end{acknowledgments}
 
\end{document}